\documentclass[a4paper,11pt]{article}
 
 \usepackage[english]{babel}  
 \usepackage[utf8]{inputenc}
 \usepackage{csquotes}
 
 \usepackage[totalwidth=17.0cm,totalheight=25.5cm]{geometry}
 \usepackage{changepage}
 
\usepackage{import}

\usepackage{amsmath,amssymb,amsthm,mathtools,mathrsfs}
\usepackage{bm,faktor}

  \newcommand{\ga}{\alpha}

  \newcommand{\gb}{\beta}

  \newcommand{\Cc}{\mathcal{C}}				
  \newcommand{\Cd}{\dot{C}}					  
  \newcommand{\Cdd}{\ddot{C}}	
  \newcommand{\Ch}{\hat{C}}

  \newcommand{\gC}{\Gamma}	
  
  \newcommand{\Dt}{\tilde{D}}				
  \newcommand{\Dh}{\hat{D}}  
  \newcommand{\gd}{\delta}					
  \newcommand{\gD}{\Delta}

  \newcommand{\eps}{\epsilon}				

  \renewcommand{\bf}{\bm{f}}				
  
  \newcommand{\fd}{\dot{f}}
  \newcommand{\fh}{\hat{f}}

  \newcommand{\Ft}{\tilde{F}}
  
  \newcommand{\Phih}{\hat{\Phi}}
  \newcommand{\Phit}{\tilde{\Phi}}
  \newcommand{\Phith}{\hat{\tilde{\Phi}}}

   \newcommand{\bgg}{\bm{g}}				

  \newcommand{\ggh}{\hat{g}}
  
  \newcommand{\ggt}{\tilde{g}}

  \newcommand{\bhd}{\dot{\bm{h}}}			
  					
  \newcommand{\hd}{\dot{h}}
  \newcommand{\hh}{\hat{h}}	
  \newcommand{\hhd}{\dot{\hat{h}}}	
  \newcommand{\bh}{\bm{h}}

  \newcommand{\It}{\tilde{I}}				
  
  \newcommand{\gi}{\iota}		
  \newcommand{\scrI}{\mathscr{I}}
  
   \newcommand{\bgk}{\bm{\kappa}} 
  \newcommand{\gk}{\kappa}
  
   \newcommand{\Kr}{\mathring{K}}
  
  \newcommand{\gL}{\Lambda}

  \newcommand{\mb}{\bar{m}}
  \newcommand{\Md}{\dot{M}}
  
  \newcommand{\mh}{\hat{m}}		
  
  \newcommand{\Mt}{\widetilde{M}}

  \newcommand{\Nd}{\dot{N}}
  				
  \newcommand{\nh}{\hat{n}}

  \newcommand{\bn}{\bm{n}}

  \newcommand{\bgO}{\bm{\Omega}}  
  \newcommand{\go}{\omega}
  \newcommand{\god}{\dot{\omega}}
  	  
  \newcommand{\goh}{\hat{\go}}		
  
  \newcommand{\gO}{\Omega}
  \newcommand{\gOh}{\hat{\Omega}}

  \newcommand{\Ph}{\hat{P}}					
  \newcommand{\Pt}{\tilde{P}}

  \newcommand{\Psit}{\tilde{\Psi}}
  
 \newcommand{\bbQ}{\mathbb{Q}}
 \newcommand{\cQ}{\mathcal{Q}}				
  
  \newcommand{\Rt}{\tilde{R}}

  \newcommand{\bbR}{\mathbb{R}}
  \newcommand{\gr}{\rho}

  \newcommand{\bgs}{\bm{\sigma}}			
  
  \newcommand{\gs}{\sigma}
  \newcommand{\gS}{\Sigma}
  \newcommand{\gsd}{\dot{\sigma}}  			
  \newcommand{\gsh}{\hat{\sigma}}

  \newcommand{\Th}{\hat{T}}					
  \newcommand{\Tt}{\widetilde{T}}

  \newcommand{\cT}{\mathcal{T}}				
  
  \newcommand{\gth}{\theta}					
  \newcommand{\gthh}{\hat{\theta}}
  \newcommand{\gTh}{\Theta}
  \newcommand{\gThh}{\hat{\Theta}}
  \newcommand{\gThd}{\dot{\Theta}}

  \newcommand{\uh}{\hat{u}}					

  \newcommand{\ud}{\dot{u}}
  \newcommand{\Ud}{\dot{U}}
   \newcommand{\udd}{\ddot{u}}
  
  \newcommand{\gU}{\Upsilon}

  \newcommand{\Vd}{\dot{V}}					

  \newcommand{\bXt}{\tilde{\bm{X}}}			

  \newcommand{\Xt}{\tilde{X}}
  \newcommand{\Xh}{\hat{X}}

  \newcommand{\bX}{\bm{X}}

\newcommand{\from}{\colon}
\newcommand{\xto}[2][]{\xrightarrow[#1]{#2}}
\newcommand{\inj}{\hookrightarrow}
\newcommand{\surj}{\twoheadrightarrow}

\DeclareMathOperator{\ISO}{ISO\!}

\newcommand{\Quotient}[2]{\faktor{#1}{#2}}

\newcommand{\covD}{\nabla}

\newcommand{\covDh}{\hat{\nabla}}
\newcommand{\covDt}{\widetilde{\nabla}}
\newcommand{\parD}{\partial}

\newcommand{\LieD}{\mathcal{L}}

\newcommand{\So}[1]{\Gamma\left[ #1 \right]}
\newcommand{\Co}[1]{\mathcal{C}^{\infty}\left( #1 \right)}

\renewcommand{\S}{\bm{\text{S}}}				

\newcommand{\Mtx}[1]{\begin{pmatrix} #1	\end{pmatrix}}
\newcommand{\sMtx}[1]{\begin{bmatrix} #1	\end{bmatrix}}  
  
\newcommand{\xeq}[1]{\stackrel{#1}{=}}

\newtheorem{Proposition}{Proposition}[section]

\newtheorem{Lemma}{Lemma}	

\theoremstyle{definition}
\newtheorem{Definition}{Definition}[section]
\numberwithin{equation}{section}

\newcommand{\MConf}{M}
\newcommand{\MRiem}{\tilde{M}}
\newcommand{\gConf}{\bm{g}}
\newcommand{\hConf}{\bm{h}}
\newcommand{\gRiem}{\tilde{g}}

\usepackage{comment}
\usepackage{enumerate}

\usepackage[backend=biber, isbn=false , url=false, hyperref=true, doi=true,giveninits=true,maxnames=4,style=numeric-comp,sorting=none]{biblatex}

\renewbibmacro{in:}{}
\DeclareFieldFormat[article]{volume}{\bibstring{volume}\addnbspace #1}
\DeclareFieldFormat[article]{number}{\bibstring{number}\addnbspace #1}
\AtEveryBibitem{\clearlist{language}\clearfield{month}}
\renewbibmacro*{journal+issuetitle}{%
	\usebibmacro{journal}%
	\setunit*{\addcomma\space}
	\iffieldundef{series}
	{}
	{\newunit
		\printfield{series}%
		\setunit{\addcomma\space}}
	\usebibmacro{volume+number+eid}%
	\setunit{\addspace}%
	\usebibmacro{issue+date}%
	\setunit{\addcolon\space}%
	\usebibmacro{issue}%
	\newunit}

\renewbibmacro*{volume+number+eid}{%
	\printfield{volume}%
	\setunit{\addcomma\space}
	\printfield{number}%
	\setunit{\addcomma\space}%
	\printfield{eid}}

\usepackage{hyperref}

\begin{document} 
	\pagestyle{plain}
	\title{Tractor geometry of asymptotically flat spacetimes}
	\author{Yannick Herfray \footnote{Yannick.Herfray@ulb.ac.be}\\ 
		{\small \it Départment de Mathématique,
			Université Libre de Bruxelles, } \\ {\small \it CP 218, Boulevard du Triomphe, B-1050 Bruxelles, Belgique. }}
	\maketitle
\begin{abstract}
In a recent work it was shown that conformal Carroll geometries are canonically equipped with a null-tractor bundle generalizing the tractor bundle of conformal geometry. We here show that in the case of the conformal boundary of an asymptotically flat spacetime of any dimension $d\geq3$, this null-tractor bundle over null infinity can be canonically derived from the interior spacetime geometry. As was previously discussed, compatible normal connections on the null-tractor bundle are not unique: We prove that they are in fact in one-to-one correspondence with the germ of the asymptotically flat spacetimes to leading order. 

In dimension $d=3$ the tractor connection invariantly encodes a choice of mass and angular momentum aspect, in dimension $d\geq4$ a choice of asymptotic shear. In dimension $d=4$ the presence of tractor curvature correspond to gravitational radiation. 

Even thought these results are by construction geometrical and coordinate invariant, we give explicit expressions in BMS coordinates for concreteness.
\end{abstract}
\maketitle

\section{Introduction}

As Geroch pointed out \cite{geroch_asymptotic_1977} one of the reasons why asymptotically flat spacetimes are such a fruitful concept in General Relativity is that they allow to make sense of an isolated system. In non-gravitational context, isolated systems are usually dealt with by choosing a ``box'' and considering the flux of whatever physical quantities is considered of interest through the boundary. The boundaries of such boxes are however always implicitly defined in terms a fixed background metric and this causes a conceptual problem for General Relativity where geometry is dynamical: the kinematic data defining the boundary and the dynamical data whose flux we are trying to measure become one same object. Geroch highlighted that asymptotically flat spacetimes precisely address this conundrum: close to null infinity some part of the geometry can be separated from the rest and taken as kinematic data while another part can be understood as dynamical and embodies the presence of gravitational radiations. We here wish to highlight that this split can be most elegantly realized in terms of tractors \cite{bailey_thomass_1994,curry_introduction_2018} and that, besides, this approach uniformly applies to asymptotically flat spacetime in any dimension $d\geq3$, shedding a clear light on how the situation differs in dimensions different than four.

If $\left(M, \ggt_{\mu\nu} = \gO^{-2} g_{\mu\nu}\right)$ is a $d$-dimensional asymptotically flat spacetime with $M = \Mt \cup \scrI$ and  $\scrI =\bbR \times S^{d-2}$, the kinematic data consist of a degenerate (conformal) metric $h_{ab} := \gi^*g_{\mu\nu}$ induced at the conformal boundary $\scrI$ together with a (weighted) vector field $n^a := g^{\mu\nu}(d\gO^{-1})_{\nu}\big|_{\scrI}$ spanning the null direction (unless otherwise specified we work with abstract indices: $\mu,\nu,...$ are indices for tensor on $M$ while $a,b,...$ are indices for tensor on $\scrI$). The resulting geometry at null infinity $\left(\scrI  , n^a, h_{ab}\right)$ has been the subject of some recent investigations under the name Carroll\footnote{The reason why these geometries deserved Lewis' Carroll name goes back to the work of Levy-Leblond \cite{levy-leblond_nouvelle_1965} on contractions of the Poincaré group: while Galilean geometries are obtained as non-relativistic limit from usual spacetimes, Carroll geometries result from taking  the opposite (sometimes called ``ultra-local'') limit. There isn't any invariant notion of times associated with Carroll geometries, rather these spacetimes can be thought as collections of spatial points with no causality relation between them. In Levy-Leblond's words, since ``absence of causality as well as arbitrarinesses in the length of time intervals is especially clear in Alice's adventures (in particular in the Mad Tea-Party) this did not seem out of place to associate L. Carroll's name''. (personal translation from \cite{levy-leblond_nouvelle_1965})} geometry  \cite{duval_carroll_2014,bergshoeff_dynamics_2014,duval_conformal_2014,duval_conformal_2014-1,bergshoeff_carroll_2017,de_boer_perfect_2018,morand_embedding_2020,ciambelli_carrollian_2019,ciambelli_carroll_2019,donnay_carrollian_2019}, see also \cite{palomo_lightlike_2021} and \cite{penrose_geometry_1972,gourgoulhon_31_2006,nurowski_intrinsic_2000,galloway_null_2004} for general literature on null hypersurfaces. Importantly, the group of symmetry of asymptotically flat spacetimes, the BMS group \cite{bondi_gravitational_1962,sachs_gravitational_1962,sachs_asymptotic_1962} (after Bondi--van der Burg--Metzner--Sachs), is also the subgroup of diffeomorphisms of $\scrI$ preserving a given \emph{conformal} Carroll geometry \cite{geroch_asymptotic_1977,duval_conformal_2014}: there is therefore a clear dictionary between asymptotic symmetries, in a neighbourhood of $\scrI$, and isomorphisms of the induced conformal Carroll geometry, which can be defined intrinsically.

The dynamical (or radiative) data in a neighbourhood of null infinity are classically understood as the appearance of asymptotic shear for a congruence of null geodesics reaching infinity: Let $\left(l^{\mu}, n^{\mu}, m^{\mu}{}_A\right)_{A\in 1...d-2}$ be a null tetrad
 such that $l^{\mu} :=\parD_r$ generates the affine parametrisation $r$ along the null geodesics of the congruence and
 \begin{align*}
 l^{\mu}n_{\mu} &=1, & m_A{}^{\mu} \mb_{B\mu} &= H_{AB},
 \end{align*}
where $H_{AB}$ is a $d-2 \times d-2$ symmetric tensor with finite limit at $\scrI$ and all other contractions vanish. The asymptotic shear $C_{AB}$ is the $d-2 \times d-2$ trace-free symmetric tensor defined as
\begin{equation}\label{Introduction: Asymptotic Shear}
 \covDt_{(\mu}l_{\nu)}\; m^{\mu}{}_{A} m^{\nu}{}_{B} \big|_0= r^{-2}\;\tfrac{1}{2} C_{AB} + O\left(r^{-3}\right).
\end{equation}
Where $\big|_0$ indicates trace-free with respect to $H_{AB}$. 

 In the physically relevant dimension $d=4$, gravitational radiations are present when all such congruences have non-vanishing asymptotic shear. The obstruction for finding a congruence whose asymptotic shear vanishes is the rescaled Weyl tensor $K^{\mu}{}_{\nu ab} n^{\nu} =  rW^{\mu}{}_{\nu a b}n^{\nu} \big|_{\scrI}$ or, equivalently, the Newman-Penrose coefficients \cite{newman_approach_1962}
 \begin{align}\label{Introduction: NP coefficients}
 \Psi_4^0 & := K_{\mu \nu \gr \gs} \; \mb^{\mu}\;n^{\nu} \; n^{\gr} \; \mb^{\gs}, & \Psi_3^0 & := K_{\mu \nu \gr \gs} \; l^{\mu}\; n^{\nu} \; n^{\gr} \; \mb^{\gs} ,  & Im\left(\Psi_2^0\right) &:= \tfrac{1}{2} K_{\mu \nu \gr \gs} \; l^{\mu}\,n^{\nu} \; m^{\gr} \; \mb^{\gs}
 \end{align}
(where $m^{\mu}$ and $\mb^{\mu}$ are now null vectors normalised as $m^{\mu} \mb_{\mu}=-1$). This result appears e.g. in \cite{ashtekar_geometry_2015} and should be thought as an asymptotic version of the Golberg-Sachs theorem \cite{goldberg_theorem_1962} (see \cite{adamo_null_2012} for a review on null geodesic congruences). It offers an elegant characterisation of gravitational radiation in terms of the behaviour of null congruences of geodesics in a neighbourhood of $\scrI$.

 How are we, however, to interpret all this from the point of view of null infinity, i.e. from the point of view of the Carroll geometry?  

A hint of the solution is given by the recent work \cite{herfray_asymptotic_2020}. In this work, it was shown that a $(d-1)$-dimensional conformal Carroll geometry  $\left(\scrI,\bn^a,\bh_{ab}\right)$ (here bold notation is used to emphasised that the fields are weighted) is \emph{canonically} associated to a $(d+1)$-dimensional vector bundle the ``null-tractor bundle''
\begin{equation*}
\cT_{\scrI} \to \scrI.
\end{equation*}
This bundle is naturally equipped with a degenerate metric $h_{IJ}$ whose kernel is spanned by a preferred section $\It^I \in \So{\cT_{\scrI}}$ (Upper latin indices $I,J,...$ will indicate abstract indices for tractors). It was however shown that connections $\Dt$ on the null-tractor bundle which are compatible with $h_{IJ}$ and $\It^I$ are not unique, not even after requiring the tractor connection to be normal (which is similar in spirit to the torsion-free condition of Riemannian geometry). We will here prove that this freedom in the choice of normal connection on the null-tractor bundle of null infinity is an invariant way of characterizing all the possible asymptotic shear \eqref{Introduction: Asymptotic Shear} of null geodesic congruences. In dimension $d=4$, the curvature of the connection is equivalent to the Newman-Penrose coefficients \eqref{Introduction: NP coefficients} and the picture is complete. In dimension $d\geq5$ the curvature will also give the obstruction for the existence of an asymptotically shear-free null congruence, this obstruction is however unrelated to presence of gravitational radiation. In $d=3$ all congruences are shear-free but there are still interesting features encoded in the tractor connection, see below.

We now discuss precisely how this equivalence can arise. It was proven in \cite{herfray_asymptotic_2020} that any choice of trivialisation $u \from \scrI \to \bbR$ for $\scrI \to S^{d-1}$ canonically defines an isomorphism
\begin{equation}\label{Introduction: Null-tractor isomorphism}
\cT_{\scrI} \xeq{u} \bbR \oplus T\scrI/n \oplus \bbR \oplus \bbR.
\end{equation}
We will re-derive this fact in the present article and relate it with the corresponding isomorphisms for
spacetime tractors. If $\Phit^I$ is a section of the null-tractor bundle we write (from now on upper latin indices $A,B,...$ are abstract indices for $T\scrI/n$)
\begin{align*}
 \Phit^I \;\xeq{u}\;\Mtx{\Phit^+ \\ \Phit^A \\ \Phit^u \\ \Phit^-}.
\end{align*}
If $\Dt$ is a normal connection on $\cT_{\scrI}$, it can then be explicitly parametrized in the above coordinates: it is a function of the data of the Carroll geometry $\left(\bn^a , \bh_{ab}\right)$ and a trace-free symmetric tensor $C_{AB}$, details will be given in proposition \ref{First order structure at null infinity: Proposition Induced Tractor connection}. On the other hand, a choice of trivialisation $u \from \scrI \to \bbR$ also defines a unique null geodesics congruence in a neighbourhood of $\scrI$ and we will prove that its asymptotic shear $C_{AB}$ matches the freedom in the tractor connection. Therefore a choice of connection on the null-tractor bundle is an invariant geometrical object at $\scrI$ whose coordinates in any trivialisation $u \from \scrI \to \bbR$ correspond to the asymptotic shear of the corresponding null geodesic congruence. It then straightforwardly follows from the interpretation of this connection as a Cartan connection (for more details on this see \cite{herfray_asymptotic_2020}) that vanishing of the curvature is equivalent to the existence of a trivialisation $u$ such that $C_{AB}=0$ (for most other choices of trivialisation $u$, it will be non-zero but however ``pure gauge'' in the precise sense that the corresponding connection is flat). For $d=4$, this gives an elegant geometrical proof that vanishing of the NP coefficients \eqref{Introduction: NP coefficients} is equivalent to the existence of an asymptotically shear-free null congruence. In dimension $d\geq5$ the situation is similar but the vanishing curvature condition is more delicate and is physically less interesting since gravitational radiations are not related to the presence of asymptotic shear.

Let us also here discuss the interesting case where $d=3$. For this singular dimension the asymptotic shear always vanish, simply because there isn't any non-zero trace-free symmetric tensor of dimension $1 \times 1$. However the null-tractor bundle still makes sense and a choice of normal tractor connection amounts to a choice of ``mass and angular momentum aspects''. The vanishing of the curvature is then equivalent to the so-called ``conservation equations'' (for a another intrinsic geometrical interpretation of the mass and angular momentum aspects as differential operators on $\scrI$, see \cite{herfray_asymptotic_2020}).\\

In \cite{herfray_asymptotic_2020}, the geometry of the null-tractor bundle of a conformal Carroll manifold $\left(\scrI,\bn^a, \bh_{ab}\right)$ and the related normal connection has been worked out in an intrinsic manner, that is without the need to refer to an asymptotically flat manifold $\left(M,\ggt_{\mu\nu} = \gO^{-2}g_{\mu\nu}\right)$ extending it. In this article we wish to explicitly show how these can be related to the geometry of the interior spacetime. Along the way we shall establish without any possible doubt that the freedom in choosing a normal connection on the null-tractor bundle and the freedom in the asymptotic shear (respectively mass and angular momentum aspects in dimension $d=3$) of the corresponding asymptotically flat spacetime are precisely the same. 

This identification will be realized by deriving the null-tractor geometry from the tractor geometry of spacetime. Let $\left(M, \ggt_{\mu\nu} = \gO^{-2}g_{\mu\nu} \right)$ be an asymptotically flat manifold, in section \ref{section: Asymptotically flat spacetimes: the conformal approach} we will first recall from \cite{curry_introduction_2018,gover_almost_2010} how this definition can be rephrased in terms of the spacetime tractors and we will, along the way, review the needed elements of tractor geometry: $\left(M, \ggt_{\mu\nu} = \gO^{-2}g_{\mu\nu}\right)$ is equivalent to a triplet $\left(M, I^{I}, \gConf_{\mu\nu}\right)$ such that $D_{\mu}I^I$ satisfies certain fall-off condition (here $D$ is the normal connection associated to the conformal metric $\gConf_{\mu\nu}$ and $I^I$ is an ``infinity tractor''). We will then show how the null-tractor bundle $\cT_{\scrI} \to \scrI$ constructed intrinsically in \cite{herfray_asymptotic_2020} can be identified with the sub-bundle $I^{\perp}$ restricted to $\scrI$. The crux of the work here, which will be achieved by the end of section \ref{section: Zeroth order structure at null infinity, the null-tractor bundle}, is to derive the corresponding transformation rules for null-tractors : because of the degeneracy of null infinity these are quite different from the usual ones. This is reflected in the fact that trivialisation $u$ (and not, as usual, choice of scales) are needed for defining the isomorphism \eqref{Introduction: Null-tractor isomorphism}. We will then show in section \ref{First order structure at null infinity, the induced tractor connection} that in the BMS coordinates given by a null geodesic congruence the normal tractor connection $D$ is asymptotically parametrized by the corresponding asymptotic shear, and that this connection induces on null-tractors a connection $\Dt$ which is itself normal in the sense of \cite{herfray_asymptotic_2020}. We will in fact prove that all such connections can be obtained in this way and that they encode the first order germ of an asymptotically flat spacetime for $d\geq4$ (respectively the second order germ for $d=3$).\\

In the context of conformal geometry, tractors and their normal connection are classical objects exposed in their modern form in \cite{bailey_thomass_1994} and going back to \cite{cartan_les_1923, thomas_conformal_1926}. That conformal geometry should play an essential role in the description of asymptotic flat spacetimes will come as no surprise, not only was it very clear from Penrose works \cite{penrose_asymptotic_1963,penrose_conformal_1964,penrose_zero_1965} but a large part of the follow up literature emphasised conformal invariance as a key features, be it the investigation of Newman H-spaces \cite{newman_heaven_1976,adamo_generalized_2010,adamo_null_2012}, the related twistor theory \cite{penrose_twistor_1973,penrose_nonlinear_1976,atiyah_twistor_2017} or Friedrich conformal equations \cite{friedrich_cauchy_1983,frauendiener_local_1999,frauendiener_conformal_2004,friedrich_peeling_2018}. In fact the realisation in \cite{friedrich_twistor_1977} that the conformally invariant local twistor transport equations \cite{penrose_twistor_1968,dighton_introduction_1974} could be described in terms of Cartan's normal connection pre-dates (and seemingly inspired) the development of tractor calculus. We will soon come back on asymptotic twistors  \cite{penrose_twistor_1973,ko_kahler_1977,mason_twistors_1985,mason_kahler_1986} which, as we shall explain, are closely related to the material presented here and in some sense anticipate on the tractor literature. Detailed investigation of tractor geometry induced at the boundary hypersurface of an asymptotically simple manifold in any dimension $d$ and its interplay with the interior geometry is however rather recent \cite{branson_conformally_2001,gover_conformal_2007,gover_almost_2010,vyatkin_manufacturing_2013,gover_boundary_2014,gover_poincare-einstein_2015,curry_introduction_2018} and mainly due to works by Gover and collaborators. We will in fact heavily draw our inspiration from these. As we shall see however, dealing with the degeneracy of the conformal Carroll geometry induced at the boundary of an asymptotically flat spacetime will imply substantial work to adapt these results (which only apply for conformal boundary which are genuine conformal manifolds). These extra efforts will be rewarded since the discrepancy between the two situations precisely encodes gravitational radiation. One illuminating comparison in this respect is the case of 2 dimensional conformal geometry where normal tractor connections are not unique but are closely related to projective complex structures \cite{calderbank_mobius_2006,burstall_conformal_2010} (see \cite{herfray_asymptotic_2020} for more on this comparison between 2D conformal and conformal Carroll geometries). Note that, in principle, the results in this article could be obtained by taking the limits from those of \cite{gover_almost_2010,curry_introduction_2018} in the limit where the cosmological constant $\gL$ goes to zero. This limit should be particularly transparent in the formalism developed in \cite{compere_lambda-bms_4_2019,compere_lambda-bms4_2020,fiorucci_charge_2021}.

Let us compare our results with the existing literature on the geometry of null infinity. The closest in spirit is probably the series of works \cite{ashtekar_radiative_1981,ashtekar_symplectic_1982,ashtekar_a._symplectic_1981,ashtekar_asymptotic_1987,ashtekar_geometry_2015,ashtekar_null_2018} which is our second main source of inspiration. In these articles, the radiative degrees of freedom at null infinity were understood as a choice of equivalence class of connections on the tangent bundle of null infinity. The relation to our work is straightforward: If $\Dt$ is a normal connection on $\cT_{\scrI}$, every choice of trivialisation $u$ for $\scrI$ will be associated with a connection $\covD$ on the tangent bundle  $T\scrI$. The set $\{\covD\}$ of all connections which can be obtained in this way then form an equivalence class of the type considered in \cite{ashtekar_radiative_1981}. In this precise sense, the equivalence classes of \cite{ashtekar_radiative_1981} are therefore equivalence classes of coordinates for an invariant geometrical object, the connection on the tractor bundle. See \cite{herfray_asymptotic_2020} for more details on this relationship. 

Other closely related works are results on asymptotic twistors  \cite{penrose_twistor_1973,ko_kahler_1977,mason_twistors_1985,mason_kahler_1986}. These are particular local twistors which are defined along $\scrI$ and they can be related to null-tractors as follows: The central result of \cite{herfray_asymptotic_2020} was to show that a $(d-1)$-dimensional conformal Carroll manifolds $\left(\scrI, \bn^a, \bh_{ab}\right)$ is equipped with a canonical $\bbR^{d-1} \rtimes \left(\bbR \times\ISO\left(d-2\right)\right)$ principal bundle $P \to \scrI$. Null-tractors are then obtained as sections of the associated bundle for the fundamental $(d+1)$-dimensional representation of this group while asymptotic twistors can be obtained as representation of the \emph{spin} group. Asymptotic twistors have, to the best of our knowledge, only be studied in dimension $d=4$ and the present article can together with \cite{herfray_asymptotic_2020} be understood as a generalisation of these to higher dimensions. In $d=4$, the geometry of asymptotic twistors is however much richer for their total space (``the'' Twistor space per say) is a 3d complex Kähler manifold, see \cite{penrose_twistor_1973,ko_kahler_1977,mason_twistors_1985,mason_kahler_1986}. We finally wish to highlight the fact that null-tractors can be treated uniformly for any dimension $d-1\geq1$ and, in this sense, serve as unifying tools for treating conformal Carroll manifolds of generic dimension.

Let us close this introduction by pointing out the recent works \cite{nguyen_effective_2021,herfray_einstein_2022} which display functionals relying on tractor methods: the first is essentially a Chern-Simon functional defined at null infinity for the tractor connection described in \cite{herfray_asymptotic_2020} while the second is a version of Einstein-Hilbert in tractor formalism.

\section{Asymptotically flat spacetimes: the conformal approach}\label{section: Asymptotically flat spacetimes: the conformal approach}

For the reader unfamiliar with this material, we here give a pedagogical introduction to conformal manifolds and tractor geometry in a form which is adapted to the investigation of asymptotically flat spacetimes. This will also allow us to set up notations and introduce the objects we shall need later on. Our main references are \cite{bailey_thomass_1994,curry_introduction_2018,gover_almost_2010}.

\subsection{Asymptotically flat spacetimes and conformal geometry}

Let $\left(\MRiem, \gRiem_{\mu\nu}\right)$ be a $d$-dimensional spacetime, $d>2$. We will use the abstract indices convention: e.g. $V^{\mu} \in \So{T\MRiem}$ is a section of the tangent bundle and $U_{\mu} \in \So{T^*\MRiem}$ a section of the dual tangent bundle.

In this article we are concerned with asymptotically flat spacetimes \cite{penrose_asymptotic_1963,penrose_conformal_1964}.

\begin{Definition}\label{Definition: Asymptotically flat spacetimes 1}
	A $d$-dimensional spacetime $\left( \MRiem, \gRiem_{\mu\nu}\right)$ is said to be \emph{asymptotically flat} (to order $k$) if there exists a spacetime $\left( \MConf, g_{\mu\nu}\right)$ with boundary $\scrI$ such that $\MRiem$ can be diffeomorphically identified with the interior, $\MRiem = \MConf\backslash \scrI$ and
	\begin{enumerate}[i)]
		\item there exists a smooth ``boundary defining function'' $\gO$ on $\MConf$, satisfying $\gO = 0$ , $d_{\mu}\gO \neq 0$ on $\scrI$ and
		\begin{equation*}
		\gRiem_{\mu\nu} = \gO^{-2} g_{\mu\nu} \quad \text{on} \quad \MRiem = \MConf\backslash \scrI,
		\end{equation*}
		\item $\gRiem$ satisfies Einstein equations $\Rt_{\mu\nu} - \frac{1}{2}\Rt \gRiem_{\mu\nu} = \Tt_{\mu\nu}$, where $\gO^{-k}\Tt_{\mu\nu}$ has a smooth limit at $\scrI$.
	\end{enumerate}
A spacetime satisfying $i)$ but not necessarily $ii)$ will be called \emph{asymptotically simple}. 
\end{Definition}
Typically one also adds to this definition constraints on the topology of the boundary (typically $\scrI = \bbR \times \S^{d-2}$ together with conditions to ensure the completeness of the boundary, see \cite{geroch_asymptotically_1978}). In most of this exposition we will however only be interested in local properties of asymptotically flat spacetimes and will not need such considerations.
 
An essential aspect of this definition is that the pair $\left(\gO,g_{\mu\nu}\right)$ is not unique: if $\left(\gO , g_{\mu\nu}\right)$ is an admissible pair then so is $\left(\go \gO, \go^{2} g_{\mu\nu} \right)$ where $\go$ is a smooth nowhere-vanishing function on $M$. We are thus really interested about the conformal class of metric $[g_{\mu\nu} \sim \go g_{\mu\nu}]$ together with the equivalence class $\left( \gO, g_{\mu\nu} \right) \sim  \left(\go \gO ,  \go^2 g_{\mu\nu}\right)$. Equivalence class of this type will be very useful and deserve some more attention.

\subsection{Conformal geometry}

A conformal manifold $\left(\MConf , \bgg_{\mu\nu} \right)$ is the data of a manifold $M$ together with an equivalence class of metric $\bgg_{\mu\nu}  = [g_{\mu\nu}]$ for the equivalence relation \begin{equation*}
g \sim \ggh \quad \Leftrightarrow \quad \ggh = \go^2 g \quad \text{where}\; \Co{M} \ni \go > 0.
\end{equation*}
The conformal class of metric $\bgg_{\mu\nu}$ can also be thought as an oriented line bundle $\cQ \subset S^2 T^*\MConf$: By construction this bundle $\cQ \to \MConf$ is such that nowhere-vanishing positively-oriented sections $g_{\mu\nu} \in \So{\cQ}$ correspond to choice of representatives $g_{\mu\nu} \in \bgg_{\mu\nu}$.

Since it admits nowhere-vanishing sections, $\cQ \to \MConf$ is always trivial. We can therefore work in a (global) trivialisation without any restriction. This is the approach that we will take throughout this paper. Accordingly, let us pick a representative $g_{\mu\nu} \in \bgg_{\mu\nu}$, then any section of $\cQ$ can be written as
\begin{equation*}
f g_{\mu\nu} \in \So{\cQ} \quad \text{where} \; f\in \Co{\MConf}.
\end{equation*}
We emphasis that $f g_{\mu\nu}$ is another ``representative'' of $\bgg_{\mu\nu}$ if and only if $f$ is everywhere positive (and in particular nowhere vanishing). 

Since we will be working in a trivialisation, all expressions will appear as functions of a metric $g_{\mu\nu} \in \So{S^2 T^*\MConf}$. Conformal invariance (or covariance) of a specific expression will then be the statement that this expression is invariant (or has a well-defined transformation rule) under the change of trivialisation
\begin{equation}\label{Asymptotically flat spacetimes: g transformation rules}
g \mapsto \ggh := \go^2 g \quad \text{where} \; \Co{\MConf} \ni \go > 0. 
\end{equation}
All our definitions will be given in terms of such transformation rules. For example we now \emph{define} for any $k \in \bbQ$ the line bundle $L^k \to \MConf$ to be such that a section $\bf \in \So{L^k}$ is given in our trivialisation by a function $f \in \Co{\MConf}$ with the transformation rule
\begin{equation*}
 f \mapsto \fh := \go^k f \quad \Leftrightarrow \quad \bf \in \So{L^k}.
\end{equation*}
Our use of bold letters for sections of $L$ is suggestive of an abstract index notation for ``weighted'' tensors. However, once again, we shall mainly be working in a trivialisation for $L$ and our use of bold letters should be restrained to the minimum.

Any other choice of conformal metric $\bh_{\mu\nu}$ can now be thought as a section of $\So{S^2T^*\MConf \otimes L^2}$, with representatives mapped from one to the other under the transformation rule
\begin{equation*}
h_{\mu\nu} \mapsto \hh_{\mu\nu} := \go^2 h_{\mu\nu} \quad \Leftrightarrow \quad \bh_{\mu\nu} \in \So{S^2T^*\MConf \otimes L^2}.
\end{equation*}

These kind of definitions can always be given a more invariant form, we will however generally refrain to do so to avoid cluttering the exposition, possibly just giving a brief outline of such definitions and pointing to other reference when they exists. For example, the $L$ bundle above can be invariantly defined as the density bundle $L := \left( \big|\gL\big|^d T^* \MConf\right)^{-\frac{1}{d}}$. In particular $L$ always exists and (contrary to what the above exposition suggests) does not rely on a choice of conformal metric for its definition. See also \cite{curry_introduction_2018} for a detailed discussion.

With this in hands, we can rephrase the first point of definition \ref{Definition: Asymptotically flat spacetimes 1} for asymptotically flat spacetimes as follows: ``\emph{there exists a conformal metric on $\MConf$, $\bgg_{\mu\nu} \in \So{\S^2T^*\MConf \otimes L^2}$, together with a choice of scale $\bgO \in \So{L}$ such that $\bgO =0$, $d\bgO\neq0$ on $\scrI$ and $\ggt_{\mu\nu}  = \bgO^{-2} \bgg_{\mu\nu} $ on $\MRiem = \MConf\backslash \scrI$}''.  As opposed to the ``physical'' $\ggt_{\mu\nu} $ which blows up at null infinity, the fields $\bgg_{\mu\nu} $ and $\bgO$ have the good property of being well-defined all over $\MConf$, they thus seems to be very natural variables to work with.

\subsection{Tractors}

The essential message from the previous subsection is that, when it comes to asymptotically flat spacetimes, working in the ``spirit'' of conformal geometry i.e. in terms of $\bgg_{\mu\nu}  \in \So{\S^2T^*\MConf \otimes L^2}$ and $\bgO \in \So{L}$ seems a lot more natural than working with the metric $\ggt_{\mu\nu}  \in \So{S^2T^*\MRiem}$. We now aim at entirely rephrasing definition \ref{Definition: Asymptotically flat spacetimes 1} solely in terms of $\bgg_{\mu\nu} $ and $\bgO$. In order to be able to rephrase the second point of definition \ref{Definition: Asymptotically flat spacetimes 1} in these terms (and in an useful way) we will however need to introduce tools from tractor calculus.

\subsubsection{The tractor bundle}\label{sssection: The tractor bundle}

Let $\left(\MConf , \bgg_{\mu\nu} \right)$ be a $d$-dimensional conformal Lorentzian manifold, i.e. of signature $(d-1,1)$. The tractor bundle $\cT \to \MConf$ is a $(d+2)$-dimensional vector bundle canonically constructed from the conformal structure. It comes with a metric of signature $(d,2)$. We here aim to give a brief and practical definition of this bundle.

In the spirit discussed in the previous subsection we will define the tractor bundle in terms of the transformation rules for its sections. The tractor bundle can however be defined more invariantly as a sub-bundle of the 2-jet of $L$, see e.g. \cite{curry_introduction_2018,bailey_thomass_1994}. One then shows that each choice of representative $g_{\mu\nu}\in \gConf_{\mu\nu}$ defines an isomorphism $\cT \xeq{g} \bbR \oplus TM \oplus \bbR$.

 Let us now come to our definition directly in terms of transformation rules. If $\Phi^I \in \So{\cT}$ is a section of the tractor bundle, we have
\begin{equation}\label{Asymptotically flat spacetimes: Tractors, transformation rules}
\Phi^I \xeq{g} \Mtx{\Phi^+ \\ \Phi^{\mu} \\ \Phi^-} \mapsto  \Phih^I \xeq{\ggh}
 \Mtx{ \go & 0 & 0 \\
 	\go^{-1}  \; \gU^{\mu} & \go^{-1}\; \gd^{\mu}{}_{\nu} & 0\\
 	-\go^{-1}\;\frac{1}{2}\gU^2 & -\go^{-1}\;\gU_{\nu}& \go^{-1}}
 \Mtx{\Phi^+ \\ \Phi^{\nu} \\ \Phi^-} \quad \Leftrightarrow \quad  \Phi^I \in \So{\cT}
\end{equation}
where $\gU_{\mu} := \go^{-1} d_{\mu} \go$ and all indices are raised and lowered with $g_{\mu\nu}$. We also define the tractor metric
\begin{equation*}
g_{IJ} := \Mtx{0 & 0 & 1 \\ 0 & g_{\mu\nu} & 0 \\ 1 & 0 & 0} \mapsto \ggh_{IJ} = \Mtx{0 & 0 & 1 \\ 0 & \go^2 g_{\mu\nu} & 0 \\ 1 & 0 & 0},
\end{equation*}
and one can check that these transformation rules are coherent i.e.
\begin{equation*}
\Phi^2 := \Phi^I g_{IJ} \Phi^I = \Phih^I \ggh_{IJ} \Phih^J.
\end{equation*}
Everywhere in this article tractor indices will be raised and lowered with the tractor metric.

An essential property of the tractor bundle is the existence of a preferred section $\bX^I \in \So{\cT \otimes L}$ defined by
\begin{equation*}
X^I = \Mtx{0 \\ 0 \\ 1} \mapsto \Xh^I = \Mtx{0 \\ 0 \\ 1}.
\end{equation*} 
The existence of this ``position tractor'' is equivalent to the fact that we have a preferred inclusion $L^{-1} \inj \cT$ and a preferred projection $\cT \surj L$
\begin{align*}
\Phi^{-} &\inj \Phi^{-} X^I = \Mtx{0 \\ 0 \\ \Phi^{-}}, & \Phi^I = \Mtx{ \Phi^+\\ \Phi^{\mu} \\ \Phi^{-}} &\surj \Phi^I X_I = \Phi^+.
\end{align*}

$\Phi^+$/$\Phi^{\mu}$/$\Phi^-$ are called primary/secondary/tertiary parts of the tractor $\Phi^I$. An important consequence is the filtration of the tractor bundle : First, remark (from the transformation rules \eqref{Asymptotically flat spacetimes: Tractors, transformation rules}) that the primary part $\Phi^+$ is a section of $L$ (this is the content of the above projection). Second, note that when the primary part vanishes then the secondary part $\Phi^{\mu}$ is a section of $T \MConf \otimes L^{-1}$. Finally when the secondary part vanished then the tertiary part $\Phi^-$ is a section of $L^{-1}$ (this is the content of the above injection).

Before we come to examples, let us stress a important point: The above definition (in terms of transformation rules) might suggest to the reader a comparison with gauge (i.e. Yang-Mills) theory. Thinking of the tractor bundle as an associated bundle for a gauge theory is however partially misleading (only partially because the tractor bundle nevertheless is an associated bundle for a Cartan geometry, see \cite{bailey_thomass_1994,friedrich_twistor_1977}) : it is not associated to an ``internal gauge symmetry'' as in Yang-Mills theory. This is more accurate to think of the tractor bundle as an extension of the tangent bundle, this extension being possible as a result of a choice of conformal metric. Just like the tangent bundle is not associated to an ``internal symmetry'' and neither is the tractor bundle. See however \cite{herfray_einstein_2022} for a presentation of the tractor bundle from a point of view which parallels the Palatini-Cartan formulation of General Relativity.

\paragraph{Example : Energy-Momentum tractor} 
Let $\left(M, \ggt_{\mu\nu}\right)$ be a pseudo-Riemannian manifold, an ``energy-momentum tensor'' is a symmetric tensor $\Tt^{\mu\nu}$ satisfying $\covDt_{\mu}\Tt^{\mu\nu} =0$. In the conformal context that we are after, it will however be more useful to consider instead trace-free tensors $T^{\mu\nu}$ and drop the differential ``conservation'' equation. For any particular scale $\ggt \in [g]$, one will then be able to recover a genuine energy-momentum tensor $\Tt^{\mu\nu} = T^{\mu\nu} + \ggt^{\mu\nu} \tfrac{1}{d}\Tt$ by solving the energy-momentum conservation $\covDt_{\nu}\left(T^{\mu\nu} \right)+ \tfrac{1}{d}\covDt^{\mu}\Tt=0$ (this fixes $T$ up to a constant).

Let $\bm{T}_{\mu}{}^{\nu}$ be a trace-free symmetric section of $\text{End}(TM) \otimes L^{-1}$. The associated Energy-Momentum tractor $T_{\mu}{}^I \in \So{T^*M \otimes \cT}$ is defined as
\begin{equation}\label{Asymptotically flat spacetimes: Energy-Momentum tractor}
	T_{\mu}{}^{I} = \Mtx{0 \\ T_{\mu} {}^{\nu} \\ -\frac{1}{d-1} \covD_{\nu} T_{\mu}{}^{\nu}}.
\end{equation}
We leave this as an exercise to the reader to check that this Energy-Momentum tractor is well defined i.e. follows the tractor transformation rules \eqref{Asymptotically flat spacetimes: Tractors, transformation rules} under
\begin{align*}
	g & \mapsto \ggh = \go^2 g, & 
	T_{\mu}{}^{\nu} &\mapsto 	\Th_{\mu}{}^{\nu}= \go^{-1} T_{\mu}{}^{\nu}.
\end{align*}
To do so, it might help to recall the transformation rules for the Levi-Civita connection :
\begin{align}\label{Asymptotically flat spacetimes: LC connection, transformation rules}
	\covD_{\mu} \ga_{\nu} &\mapsto \covDh_{\mu} \ga_{\nu} =  \covD_{\mu} \ga_{\nu} -\gU_{\mu} \ga_{\nu} -\ga_{\mu} \gU_{\nu} + \gU^{\gr} \ga_{\gr} g_{\mu\nu},\\	
		\covD_{\mu} \xi^{\nu} &\mapsto \covDh_{\mu} \xi^{\nu}  =\covD_{\mu} \xi^{\nu} + \gU_{\mu} \xi^{\nu} - \xi_{\mu} \gU^{\nu} + \xi^{\gr} \gU_{\gr} \;\gd^{\mu}_{\nu}.\nonumber
\end{align}

\paragraph{Example : Infinity tractors and Thomas operator}

A crucial (and in fact defining), property of the tractor bundle is that it comes equipped with a preferred differential operator $I \from \So{L} \to \So{\cT}$, the so-called Thomas operator:
\begin{equation}\label{Asymptotically flat spacetimes: Thomas operator, Definition}
	I\left|\begin{array}{ccc}
		\So{L} & \to & \So{\cT} \\
		\sigma & \mapsto & I(\gs)^I = \Mtx{\gs\\ \covD^{\mu} \gs \\ -\frac{1}{d}( \Delta\gs + P \gs) }
	\end{array}\right.
\end{equation}
where $\Delta = g^{\mu\nu}\nabla_{\nu} \nabla_\mu$ and $P := \frac{1}{2(d-1)}R$ (with $R$ the scalar curvature of $g_{\mu\nu}$).  This an instructive exercise to check that this operator is well defined i.e. follows the tractor transformation rules \eqref{Asymptotically flat spacetimes: Tractors, transformation rules} under
\begin{align*}
	g & \mapsto \ggh = \go^2 g, & 
	\gs &\mapsto \gsh = \go \gs.
\end{align*}
This can be done explicitly making use of the transformation rules \eqref{Asymptotically flat spacetimes: LC connection, transformation rules} for the Levi-Civita connection and those for the scalar curvature,
\begin{align*}
P&\mapsto \Ph = \go^{-2} (P -\nabla^{\gr} \gU_{\gr} - \frac{d-2}{2} \gU^2).
\end{align*}
We will call infinity tractors, tractors $I^I \in \So{\cT}$ which are in the image of Thomas operator $I^I = I(\gs)^I$ .

\subsubsection{The normal tractor connection}

The essential reason why the tractor bundle is interesting is the existence of a preferred metric-preserving connection, the normal tractor connection. Here ``normal'' refers to a constraint on the curvature that needs to be imposed on the connection to obtain unicity. This is similar to the situation in Riemannian geometry: there are many metric-preserving connection on the tangent bundle but a unique torsion-free connection, the Levi-Civita connection. In conformal geometry there are many metric-preserving connection on the tractor bundle but a unique ``normal'' connection. Once again, and as we already pointed out in the previous subsection, this is a useful point of view to think of the tractor bundle as a generalisation of the tangent bundle suited to conformal geometry.
In our philosophy of making this presentation as straight to the point as possible we will not discuss how to state the normality conditions but simply give the ``final answer'' i.e. the explicit form of the normal tractor connection, see \cite{bailey_thomass_1994,curry_introduction_2018,herfray_einstein_2022} for more details.

In order to give the explicit form of the normal tractor connection, we need to recall the definition of the Schouten tensor and its trace:
\begin{align*}
P_{\mu\nu} &:= \frac{1}{d-2}\left(R_{\mu\nu} -\frac{R}{2(d-1)}g_{\mu\nu}, \right), & P :=\frac{1}{2(d-1)}R 
\end{align*}
where $R_{\mu\nu}$ is the Ricci tensor and $R$ the Ricci scalar. It will also be useful to have the transformation rules for the Schouten tensor
\begin{equation}\label{Asymptotically flat spacetimes: Schouten tensor, transformation rules}
P_{\mu\nu} \mapsto \Ph_{\mu\nu} = P_{\mu\nu} - \nabla_{\mu} \gU_{\nu} + \gU_{\mu}\gU_{\nu} -\frac{1}{2}\gU^2 g_{\mu\nu}.
\end{equation}

Armed with these remarks, we define the normal tractor connection $D$ through the relation
\begin{equation}\label{Asymptotically flat spacetimes: Tractor connection, Definition}
 D_{\gr} \Phi^I := \Mtx{\covD_{\gr} & -g_{\gr \nu} & 0 \\
 	 P^{\mu}{}_{\gr} & \gd^{\mu}{}_{\nu}\nabla_{\gr} &\gd^{\mu}{}_{\gr} \\
 0 & - P_{\gr \nu} & \covD_{\gr}
  } \Mtx{\Phi^+ \\ \Phi^{\nu} \\ \Phi^-}.
\end{equation}
Making use of the transformation rules \eqref{Asymptotically flat spacetimes: g transformation rules},\eqref{Asymptotically flat spacetimes: Tractors, transformation rules},\eqref{Asymptotically flat spacetimes: Schouten tensor, transformation rules} and \eqref{Asymptotically flat spacetimes: LC connection, transformation rules} one can can check that $\Dh \Phih^I = \hat{D \Phi^I}$ and that this connection is indeed well-defined.

A direct computation then shows that the tractor curvature is
\begin{equation}\label{Asymptotically flat spacetimes: Tractor curvature, Definition}
F^I{}_J{}_{\mu\nu} = \Mtx{0 &0 &0 \\
	C_{\mu\nu}{}^{\gr} & W^{\gr}{}_{\gs \mu\nu} & 0 \\
	0 & -C_{\mu\nu\gs} &0
   }
\end{equation}
where $W^{\mu}{}_{\nu\gr\gs}$ and $C_{\mu\nu}{}^{\gr}$ respectively stand for the Weyl and the Cotton tensor
\begin{align*}
	W^{\mu}{}_{\nu\gr\gs} & := R^{\mu}{}_{\nu\gr\gs} - 2 P^{\mu}{}_{[\gr} g_{\gs] \nu} - 2 \gd^{\mu}{}_{[\gr} P_{\gs ]\nu}, & C_{\mu\nu}{}^{\gr} &:= 2 \covD_{[\mu} P_{\nu]}{}^{\gr}.
\end{align*}

In dimension $d>3$, the tractor curvature vanishes if and only if the Weyl curvature vanishes. In dimension $d=3$ it vanishes if and only if the Cotton tensor vanishes. This two facts are direct consequences of the first of the identities
\begin{align}\label{Asymptotically flat spacetimes: Cotton tensor, identities}
	\left(d-3\right) C_{\mu\nu}{}^{\gr} &= \covD_{\gs} W^{\gs \gr}{}_{\mu\nu}, &
	C_{\gr \nu}{}^{\gr} &= \covD_{\gr} P_{\nu}{}^{\gr} - \covD_{\nu} P =0.
\end{align}
These two relations can themselves be derived from Bianchi identity $\covD_{[\mu} R_{\nu \gr]\gs \eta} =0$.

\paragraph{Example : Infinity tractors and Energy-momentum tractors}

It is an enlightening exercise to check the following facts. First, a generic tractor $I^I= \Mtx{ \gs, & I^{\mu}, & I^+}$ is an infinity tractor $I^I = I\left(\gs\right)^I$ if and only if it satisfies
\begin{align*}
	D_{\gr}I^+ &=0 & D_{\mu}I^{\mu}&=0.
\end{align*}
Second, let $T_{\mu}{}^I = \Mtx{ 0,& T_{\mu}{}^{\nu}, & T_{\mu}{}^-}$ where $T_{\mu}{}^{\nu}$ is a trace-free symmetric tensor (note that $D_{\gr}I\left(\gs\right)^I$ is always of this form by the previous remark). The exterior derivative $D_{[\mu} T_{\nu]}{}^I$ of such fields automatically satisfies $D_{[\mu} T_{\nu]}{}^+=0$. What is more, $T_{\mu}{}^I$ is an energy-momentum tractor (i.e. is of the form \eqref{Asymptotically flat spacetimes: Energy-Momentum tractor}) if and only if $D_{[\mu} T_{\nu]}{}^{\mu}=0$.

Putting these two results together one easily derives that the covariant derivative of an infinity tractor $D_{\mu}I\left(\gs\right)^I$ always is an energy-momentum tractor (this is because $D_{[\mu} D_{\nu]}I\left(\gs\right){}^{\mu} = F^{\mu}{}_{J\mu\nu}I^J$ which can be seen to vanish by normality). In other terms we always have
\begin{equation*}
	D_{\mu}I\left(\gs\right)^I = T_{\mu}{}^I
\end{equation*}
where $T_{\mu}{}^I$ is of the form \eqref{Asymptotically flat spacetimes: Energy-Momentum tractor}.

\subsection{Almost Einstein manifolds}

\subsubsection{Einstein Equations}

Let $\left(\MConf , \bgg_{\mu\nu}\right)$ be a conformal manifold and let $D$ be the associated normal tractor connection (defined for a given representative $g_{\mu\nu}$ by eq \eqref{Asymptotically flat spacetimes: Tractor connection, Definition}). Recall that if $\bgO \in \So{L}$ is a nowhere-vanishing scale then $\gO^{-2} g_{\mu\nu}$ is a genuine metric. We note $I\left(\gO\right)^I$ the image of $\bgO$ by Thomas operator (defined for a fixed representative $g_{\mu\nu}$ by \eqref{Asymptotically flat spacetimes: Thomas operator, Definition}).

 One reason why tractor calculus is well suited for studying asymptotically flat spacetimes is that Einstein equations take an especially convenient form:
\begin{Proposition}\label{Proposition: tractor Einstein equations}
Let $\left(\MConf , \bgg_{\mu\nu}\right)$ be a conformal manifold. There exists a representative $\ggt_{\mu\nu} := \gO^{-2} g_{\mu\nu}$ which satisfies Einstein vacuum equations if and only if there exists a covariantly constant section of the tractor bundle \begin{equation*}
	D_{\gr}I^I =0, \; I^I \in \So{\cT}
	\end{equation*} such that $I^I \bX_I \in \So{L}$ is nowhere vanishing. 
	
	Then $\bgO = I^I \bX_I$, $I^I$ is an infinity tractor $I^I = I(\gO)^I$ and the scalar curvature of $\ggt_{\mu\nu}$ is given by $\Rt = - d(d-1) I^2$.
\end{Proposition}

Before we discuss the proof of this standard result a few remarks are in order. First, we emphasise that the infinity tractor selects a preferred metric $\ggt_{\mu\nu} := \gO^{-2} g_{\mu\nu}$ out of the conformal class $\bgg_{\mu\nu}$ and in this sense ``breaks conformal invariance'' (This is similar to the infinity twistor of twistor theory; in fact, in $d=4$, local twistors \cite{penrose_twistor_1973} are spin extensions of tractors \cite{friedrich_twistor_1977} and the above result can be found in local twistor notation in \cite{frauendiener_local_1999} with $I^I$ then corresponding to the infinity twistor $I^{\ga\gb}$). A second important remark is that by ``Einstein vacuum equations'' we here mean vanishing of trace-free Ricci tensor $\Rt_{\mu\nu}\big|_0 =0$ : Bianchi identities \eqref{Asymptotically flat spacetimes: Cotton tensor, identities} then imply that $\covD_{\mu}\Rt=0$ i.e. scalar curvature is constant but its precise value is not fixed. Thus the reader might want to include $I^2 = -\gL$ as an extra equation in the proposition to differentiate possible cosmological constants.

The proof of the above proposition is a good exercise in tractor calculus and what is more, involves some partial results that will be immediately useful for us. For this reason, we now give a partial proof. See \cite{curry_introduction_2018} for a complete discussion.

\begin{proof}
 Let $I^I$ be a section of the tractor bundle such that $\bgO := I^I\bX_I$ is nowhere vanishing and consider the equations
\begin{equation*}
D_{\gr} I^I = \Mtx{\parD_{\gr} & -g_{\gr \nu} & 0 \\
	P^{\mu}{}_{\gr} & \gd^{\mu}{}_{\nu}\covD_{\gr} &\gd^{\mu}{}_{\gr} \\
	0 & - P_{\gr \nu} & \parD_{\gr}
} \Mtx{ \gO \\ I^{\nu} \\ I^-} =0.
\end{equation*}
Solving for the first line and the trace of the second one finds that $I^I= I\left(\gO\right)$ i.e. $I^I$ must be the image of $\gO$ by Thomas operator. It then follows that $DI^I =0$ is equivalent to the vanishing of
\begin{equation*}
D_{\gr} \big(I\left(\gO\right)\big) ^I =  
\Mtx{\parD_{\gr} & -g_{\gr \nu} & 0 \\
	P^{\mu}{}_{\gr} & \gd^{\mu}{}_{\nu}\covD_{\gr} &\gd^{\mu}{}_{\gr} \\
	0 & - P_{\gr \nu} & \parD_{\gr}
} \Mtx{ \gO \\ \covD^{\nu}\gO\\ -\frac{1}{d}\left(\gD\gO + P\gO\right) }.
\end{equation*}
Noting with a tilde all tensors constructed from $\ggt_{\mu\nu} := \gO^{-2} g_{\mu\nu}$, one can prove that 
\begin{align}
	D_{\gr} \big(I\left(\gO\right)\big)^{+}  &= 0,\\
	D_{\gr} \big(I\left(\gO\right)\big)^{\mu} &=  \gO\frac{1}{(d-2)} g^{\mu\nu}\Rt_{\nu}{}_{\gr}\big|_{0},\label{Asymptotically flat spacetimes: tractor Einstein equations trace-free}\\
	D_{\gr} \big(I\left(\gO\right)\big)^{-}  &=  -\gO^{-1} \frac{1}{2d(d-1)} \parD_{\gr} \Rt - \frac{1}{d-2}\parD^{\nu}\gO \;\Rt_{\nu}{}_{\gr}\big|_{0}, \label{Asymptotically flat spacetimes: tractor Einstein equations trace}	
\end{align}
where $\big|_{0}$ stands for ``trace-free part of''.

 As was proved in the example at the end of the previous section, $D_{\gr}I\left(\gs\right)^I$ always has the form of an energy momentum tensor \eqref{Asymptotically flat spacetimes: Energy-Momentum tractor} and is therefore zero if and only if its secondary part vanishes. One sees from \eqref{Asymptotically flat spacetimes: tractor Einstein equations trace-free} that this is equivalent to Einstein vacuum equations for $\ggt_{\mu\nu}$. This proves one direction of the equivalence. However, if $\gO^{-2}g_{\mu\nu}$ is Einstein, the same reasoning shows that $D \big(I\left(\gO\right)\big) ^I = 0$. Finally a direct computation shows that $I^2\left(\gO\right)  = -\frac{2}{d}\Pt$. 
 
Note that whenever $\gO$ is nowhere vanishing, one can make use of the transformation rule $\gO \mapsto \go \gO$ with $\So{M} \ni \go>0$ to achieve $\gO=1$ i.e. $\ggt_{\mu\nu}= g_{\mu\nu}$. Making use of this gauge fixing would have given a straightforward proof. However, in what follows we will be interested in situation where $\gO$ vanishes at a certain locus and this will be convenient to have a proof not relying on this gauge fixing. 
 \end{proof}
 
\subsubsection{Almost Einstein manifolds}

Proposition \ref{Proposition: tractor Einstein equations} asserts that vacuum solutions of Einstein equations are equivalent to a pair $\left(\bgg_{\mu\nu} , I^I\right )$ such that $\gO := I^I X_I$ is nowhere vanishing and $D_{\gr}I^I =0$. An essential remark is that requiring $\gO := I^I X_I$ to be nowhere vanishing is only necessary for interpreting $\gO^{-2}g$ as a metric (since this last object is not defined at points where $\gO=0$), however the equation $D_{\gr}I^I=0$ is well-defined even in spacetime regions where $\gO =0$. This suggests to introduce (following \cite{gover_almost_2010,curry_introduction_2018}) \emph{ almost Einstein manifolds} as pairs $\left(\bgg_{\mu\nu} , I^I\right )$ with $D_{\gr}I^I=0$, now allowing for $\gO :=I^IX_I$ to vanish on a hyper-surface. By definition almost Einstein manifolds $\left(\MConf , \bgg_{\mu\nu} , I^I\right )$ are such that at any spacetime points ``in the interior'' i.e. such that $\gO := I^IX_I \neq 0$ the metric $\gO^{-2}g_{\mu\nu}$ satisfies Einstein vacuum equations while at spacetimes points ``at infinity'', i.e. such that $\gO=0$, the metric is ill-defined. Note however that both the conformal metric $\bgg_{\mu\nu}$ and the infinity tractor $I^I$ are well-defined everywhere, including points ``at infinity''.

All of this suggests to reformulate asymptotically flat spacetimes as a weakening of almost Einstein manifolds:
\begin{Definition}\label{Definition: Asymptotically flat spacetimes 2}
	Let  $\left( \MConf  , \bgg_{\mu\nu} \right)$ be a conformal manifolds with boundary $\scrI$ and $I^I \in \So{\cT}$ a section of the tractor bundle, we will say that  $\left(\MConf , \bgg_{\mu\nu} , I^I   \right)$  defines an asymptotically flat spacetime (to order $k$) if and only if
	\begin{enumerate}[i)]
		\item $\gO := I^I X_I$ vanishes at $\scrI := \parD \MConf$ only,
		\item $D_{\gr} I^I  = T_{\gr}{}^I$ where $T_{\gr}{}^I$ is of the form $T_{\gr}{}^I=\Mtx{0,& T_{\mu}{}^{\nu}, & T_{\mu}{}^+}$ with $T_{\mu}{}^{\mu}=0$ and the rescaled tensor $\gO^{-(k+1)} T_{\gr}{}^{\mu}$ has a well-defined smooth limit at $\scrI$, 
		\item $I^2 =0$ at $\scrI$.
	\end{enumerate}
\end{Definition}
Here, $\text{iii})$ is only needed to discriminate possible cosmological constants (see also the discussion after proposition \ref{Proposition: tractor Einstein equations}). In fact the above definition applies uniformly to asymptotically AdS/flat/dS spacetime by taking $I^2 =-\gL$, were $\gL$ is the cosmological constant. The rest of the definition is justified by the following.
\begin{Proposition}\label{Proposition: Asymptotically flat spacetimes 2}
Asymptotically flat spacetimes $\left(\MRiem , \ggt_{\mu\nu} \right)$ in the sense of definition \ref{Definition: Asymptotically flat spacetimes 1} are in one-to-one correspondence with asymptotically flat spacetimes $\left(\MConf , \bgg_{\mu\nu} , I^I   \right)$ in the sense of definition \ref{Definition: Asymptotically flat spacetimes 2}.

 One has $I^{I}= I(\gO)^I$, $\ggt_{\mu\nu}= \gO^{-2}g_{\mu\nu}$ and $\tfrac{\gO}{d-2}\Tt_{\mu\nu}\big|_0 = T_{\mu\nu}$.
\end{Proposition}

From the tractor perspective it is slightly more natural to require a fall-off on the energy-momentum tractor than on the energy-momentum tensor. To accommodate this, one needs to make a minor change to the definition of asymptotically flat spacetime:
\begin{Proposition}\label{Proposition: Asymptotically flat spacetimes ALT}
If we replace the second point in definition \ref{Definition: Asymptotically flat spacetimes 2} by
	\begin{enumerate}
\item[ii)] $D_{\gr} I^I  = T_{\gr}{}^I$ where $T_{\gr}{}^I$ is of the form $T_{\gr}{}^I=\Mtx{0,& T_{\mu}{}^{\nu}, & T_{\mu}{}^+}$ with $T_{\mu}{}^{\mu}=0$ and the rescaled tractor $\gO^{-(k+1)} T_{\gr}{}^{I}$ has a well-defined smooth limit at $\scrI$, 
\end{enumerate}
then the resulting spacetime is asymptotically flat (to order $k$) together with the extra requirement that $\gO^{-(k+3)}\Tt_{\mu\nu} \ggt^{\mu\nu}$ must have a smooth limit at $\scrI$ (i.e. the trace of $\Tt_{\mu\nu}$ must vanish one order faster than required in definition \ref{Definition: Asymptotically flat spacetimes 1}).
\end{Proposition}
It might be useful to remark that the condition in the above proposition is equivalent to requiring that $I^I= I\left(\gO\right)^I$ and $\gO^{-(k+1)}D_{\gr}I\left(\gO\right)^I$ has a smooth limit. These propositions follows from the previous discussion and are direct generalisations of proposition \ref{Proposition: tractor Einstein equations}.

\begin{proof}
We first concentrate on proposition \ref{Proposition: Asymptotically flat spacetimes 2}.
 One direction is straightforward: If $\left(\MRiem , \ggt_{\mu\nu}\right)$ is asymptotically flat in the sense of definition \ref{Definition: Asymptotically flat spacetimes 1} then it uniquely defines a triplet $\left( \MConf , \bgg_{\mu\nu} , I(\bgO)^I \right)$ which is asymptotically flat in the sense of definition \ref{Definition: Asymptotically flat spacetimes 2} (this will be clear from what follows).

 To see the converse, first note that, under the hypothesis of the proposition, $D_{\gr}I^I = T_{\gr}{}^I$ implies $D_{\gr}I^+=0$ and $D_{\mu}I^{\mu}=0$. As was discussed at the end of the previous subsection, these last two equations are in fact equivalent to $I^{I} = I(\gO)^I$. Since $\gO := I^I X_I$ is supposed to be nowhere vanishing in the interior $\MRiem$ of $\MConf$ this defines a pseudo-Riemannian metric $\left(\MRiem , \ggt := \gO^{-2}g_{\mu\nu}\right)$. As we previously remarked $D_{\gr}I(\gO)^I$ is then automatically an energy-momentum tractor. Consequently, $D_{\gr}I^I = T_{\gr}{}^I$ is in fact equivalent to
\begin{align}\label{Asymptotically flat spacetimes: proof of equivalence 1) }
D_{\gr}I(\gO)^{\mu}&= T_{\gr}{}^{\mu} & D_{\gr}I(\gO)^{-}&= -\tfrac{1}{d-1}\covD_{\nu}T_{\gr}{}^{\nu}. 
\end{align}
Which, from equation \eqref{Asymptotically flat spacetimes: tractor Einstein equations trace-free},\eqref{Asymptotically flat spacetimes: tractor Einstein equations trace}, can be rewritten as
\begin{align*}
	\tfrac{1}{d-2}\Rt_{\mu\nu}\big|_0 &= \gO^{-1} T_{\mu\nu}, & 		-\tfrac{1}{2d(d-1)}\parD_{\gr} \Rt &= -\tfrac{\gO }{d-1}\covD_{\mu} T_{\gr}{}^{\mu} + \tfrac{1}{d-2} T_{\gr\mu}\parD^{\mu}\gO.
\end{align*}
Since by hypothesis $\gO^{-(k+1)} T_{\mu\nu}$ has a well-defined limit at $\scrI$ and $\Rt\big|_{\scrI} = -d(d-1)I^2\big|_{\scrI}$ vanishes, these last equations imply that both $\gO^{-k}\Rt_{\mu\nu}\big|_0$ and $\gO^{-(k+2)}\Rt$ must have a well-defined limit at $\scrI$. This concludes the proof of proposition \ref{Proposition: Asymptotically flat spacetimes 2}.

If we consider the strongest fall-off condition of proposition \ref{Proposition: Asymptotically flat spacetimes ALT} then $D_{\gr}I^I = T_{\gr}{}^I$ is found to be equivalent to 
\begin{align*}
	\tfrac{1}{d-2}\Rt_{\mu\nu}\big|_0 &= \gO^{-1} T_{\mu\nu}, & 		-\tfrac{1}{2d(d-1)}\parD_{\gr} \Rt &= \gO  T_{\gr}{}^- + \tfrac{1}{d-2} T_{\gr\mu}\parD^{\mu}\gO.
\end{align*}
where both $\gO^{-(k+1)} T_{\mu\nu}$ and $\gO^{-(k+1)} T_{\mu}{}^-$ have smooth limits. This is equivalent to $\gO^{-k} \Rt_{\mu\nu}\big|_0$ and $\gO^{-(k+3)} \Rt$ having smooth limits at $\scrI$.
\end{proof}

As far as the author is aware propositions \ref{Proposition: Asymptotically flat spacetimes 2} and \ref{Proposition: Asymptotically flat spacetimes ALT} were first stated (or rather clearly hinted at) in \cite{curry_introduction_2018} with the essential idea of ``almost Einstein manifold'' however going back to \cite{gover_almost_2005,gover_almost_2010}. It follows that definition \ref{Definition: Asymptotically flat spacetimes 2} could perfectly be taken as an alternative definition for asymptotically flat spacetimes - this is essentially the philosophy that we will pursue in the rest of this article. 

We wish to stress that from the point of view of the conformal geometry, not only all fields (i.e. both the conformal metric $\bgg_{\mu\nu}$ and the infinity tractor $I^I$) but also the field equations $D_{\gr}I^I=0$ are well-behaved everywhere on $\MConf$ (including at $\scrI$). This makes studying asymptotically flat spacetimes from this point of view especially appealing (and indeed this is a version of this idea which underlies Friedrich conformal equations which were very fruitful in making progress on the global problem, see \cite{friedrich_cauchy_1983,frauendiener_local_1999,frauendiener_conformal_2004,friedrich_peeling_2018}). One motivation for this article is to show how fruitful this point of view is by giving an elegant description of the geometry induced at null infinity.  As we will now discuss, once working in a conformally covariant manner the relationship between the induced geometry at $\scrI$ and the interior geometry is completely transparent and reasonably straightforward.

\paragraph{Differentiability}\mbox{}

In this context the amount of differentiability on $\left( \gConf_{\mu\nu} , \bgO \right)$ that one is willing to require at  $\scrI$ has been at the heart of numerous discussions (see \cite{chrusciel_gravitational_1995, friedrich_peeling_2018} and reference therein). In what follows, we will require that $\left( \gConf_{\mu\nu} , \bgO \right)$ is of class $\Cc^{3}$. This is because we will only need to suppose that
\begin{equation*}
	DI\left(\gO\right)^I = O\left(\gO^2\right),
\end{equation*}
Here and everywhere in the rest of this article $O\left(\gO^k\right)$ will indicate a function $f$ in a neighbourhood of $\scrI$ such that the restriction of $\gO^{-k}f$ on $\scrI$ is a well-defined smooth function.

Inspections shows that one is not fully using these requirements and that class $\Cc^{2}$ and $DI\left(\gO\right)^I = o\left(\gO\right)$ would be sufficient. These last differentiability requirements are still strong enough to fit in the class of poly-homogeneous spacetimes that have a finite shear as discussed e.g. in \cite{kroon_conserved_1998,godazgar_bms_2020} but would not imply the peeling for example.

\section{Zeroth order structure at null infinity, the null-tractor bundle}\label{section: Zeroth order structure at null infinity, the null-tractor bundle}

From now-on we will always assume that asymptotically flat spacetimes $\left(\MConf , \gConf_{\mu\nu} , \bgO\right)$ we consider satisfy
\begin{equation*}
D_{\gr}I\left(\gO\right)^I = O\left(\gO^2\right).
\end{equation*} 
By proposition \ref{Proposition: Asymptotically flat spacetimes ALT} this amounts to requiring that the physical metric satisfies $\Rt_{\mu\nu}\big|_0 = O\left(\gO\right)$, $\Rt = O\left(\gO^4\right)$. 

We call ``zeroth'' order structure the geometrical structure induced at $\scrI$ by restriction of $g_{\mu\nu}$ and $g^{\mu\nu}(d\gO)_{\nu}$. The resulting geometry at null infinity is the data $\left(\scrI,\hConf_{ab}, \bn^a\right)$ of a degenerate conformal metric $\hConf_{ab}$ together with a weighted vector field $\bn^a$ spanning its kernel (Here and everywhere, our convention is to use small latin indices $a,b,...$ as abstract indices for tensors on $\scrI$). This essentially correspond (in the four dimensional context, $d=4$) to the ``universal structure'' from \cite{geroch_asymptotic_1977,ashtekar_a._symplectic_1981,ashtekar_asymptotic_1987,ashtekar_geometry_2015,ashtekar_null_2018}. This is also a conformal version of the ``weak'' Carroll structure from \cite{levy-leblond_nouvelle_1965,duval_conformal_2014,duval_conformal_2014-1,morand_embedding_2020,ciambelli_carroll_2019}. Following this recent literature we will call this induced data a conformal Carroll geometry

It was shown in \cite{herfray_asymptotic_2020} that this rather elementary geometrical structure is enough to be able to define a null-tractor bundle $\cT_{\scrI}$ at $\scrI$ with property essentially similar to the tractor bundle $\cT$ of $M$. In this section we will prove that this intrinsic tractor bundle at $\scrI$ can also be canonically identified with the restriction of (a sub-bundle of) the spacetime tractor bundle. Particular care is given to the definition of the splitting isomorphisms $\cT_{\scrI} \xeq{u} \bbR \oplus T\scrI/n \oplus \bbR \oplus \bbR$ and corresponding transformation rules.  The crux of this section will be to prove that the whole construction is indeed intrinsic, i.e. does not depend of the details of the chosen extension $\left(M , \gConf_{\mu\nu} , \bgO\right)$ but only on the boundary conformal Carroll geometry $\left(\scrI , \hConf_{ab} , \bn^a \right)$. For convenience, the main results concerning null-tractors are summarised by the end of this section. 

At the end of this section we also discuss the more invariant definition of null-tractors in terms of the second jet bundle $J^2L_{\scrI}$ at $\scrI$.

\subsection{Conformal geometry of null infinity}

\subsubsection{Conformal Carroll geometry}

 Since $\left(M, \bgg_{\mu\nu}\right)$ is taken to be a genuine conformal manifold one can simply restrict the conformal metric at the boundary $\scrI= \parD M$. If $\gi$ is the inclusion $\scrI \xto{\gi} M$, the induced conformal metric is
\begin{equation*}
\bh_{ab} := \gi^*\bgg_{\mu\nu}  \in \So{S^2 T^*\scrI \otimes (L_{\scrI})^2}
\end{equation*}
The resulting tensor is a section of $S^2 T^*\scrI \otimes (L_{\scrI})^2$, which will practically means that it can be represented by a symmetric tensor $h_{ab} \in \So{S^2 T^*\scrI}$ with transformations rules
	\begin{equation*}
	h_{ab} \mapsto \hh_{ab}= (\go_{0})^2 \; h_{ab} 
	\end{equation*}
where $\go_0 := \go \big|_{\scrI}$. 

It follows from the definition of asymptotically flat spacetimes and Thomas operator that 
\begin{equation*}
I(\gO)^I \big|_{\scrI}  = \Mtx{0 \\ n^{\mu} \\ -\frac{1}{d} \covD_{\mu} n^{\mu}}.
\end{equation*}
Where $n^{\mu} := g^{\mu\nu} d_{\nu}\gO\big|_{\scrI}  $ is the normal at $\scrI$. One has the transformation rules
\begin{align*}
n^{\mu} & \mapsto  (\go_{0})^{-1} n^{\mu}.
\end{align*}

By proposition \ref{Proposition: Asymptotically flat spacetimes 2} the definition of asymptotically flat spacetimes implies that
 \begin{equation*}
	0 = I^2\big|_{\scrI} = n^2\big|_{\scrI}.
\end{equation*} One therefore recover the well-known fact that the conformal boundary of an asymptotically flat spacetimes is null. In particular, the normal is tangential at $\scrI$ and we will therefore write it indifferently as $n^a$ or $n^{\mu}$. Together with the transformation rules for $n^{\mu}$, this implies that the normal really is a weighted section of tangent bundle at $\scrI$,
 \begin{equation*}
 	\bn^a \in \So{T\scrI \otimes (L_{\scrI})^{-1}}.
 \end{equation*} 
  Since tangent vectors to $\scrI$ necessarily have a zero inner product with the normal, we have
\begin{equation*}
n^b h_{ab} =0
\end{equation*}
i.e. $h_{ab}$ is a degenerate metric whose kernel is spanned by $n^a$. 
Taken together the induced conformal metric $\bh_{ab}$ and the normal $\bn^a$ form a conformal Carroll geometry (from \cite{duval_carroll_2014,duval_conformal_2014-1}). 
\begin{Definition}
A conformal Carroll geometry $\left(\scrI , \bh_{ab} , \bn^a\right)$ on a $(d-1)$-dimensional manifold $\scrI$ is the data of a nowhere-vanishing weighted vector field $\bn^a \in \So{T\scrI \otimes (L_{\scrI})^{-1}}$ and a non-invertible symmetric tensor $\hConf_{ab} \in \So{S^2 T^*\scrI \otimes (L_{\scrI})^2}$ whose kernel is generated by $\bn^a$.
\end{Definition}
From now on, we will also suppose that the quotient of $\scrI$ by the integral lines of $\bn^a$ is a smooth $(d-2)$-dimensional manifold $\gS$ and that $\scrI$ is the total space of a trivial fibre bundle $\scrI \to \gS$. This si simply for convenience since all our results will be local. 

\subsubsection{The quotient tangent bundle and Ehresmann connections}\label{sssection: The quotient tangent bundle and Ehresmann connections}

At this stage we still haven't used the condition $DI(\gO)^I\big|_{\scrI}=0$. Before we come to this, it will be useful to introduce some notations.

\paragraph{The tangent bundle mod $n$.}

If $X^a$,$Y^a \in T_x\scrI$ are tangent vectors at a point $x$ of $\scrI$, one introduce the equivalence relation $X^a \sim Y^a$ if and only if $X^a - Y^a = f n^a$ for some $f\in \bbR$. We write $T_x\scrI/n := \{ [X^a] , X^a \in T_x\scrI \} $. The tangent bundle mod $n$, $T\scrI/n := \cup_{x \in \scrI} T_x\scrI/n $, is then a smooth vector bundle of rank $n-2$ on $\scrI$. 

Let us introduce a bit more notations for tensors at $\scrI$. As already discussed we use \emph{lower} case Latin letter from the beginning of the alphabet to represent tensor indices associated to the tangent bundle of $\scrI$, e.g. $\bn^a \in \So{ T\scrI \otimes L^{-1}_{\scrI}}$, $\hConf_{ab} \in \So{S^2(T\scrI)^* \otimes L^{2}_{\scrI}}$.  We then use \emph{upper} case Latin letter from the beginning of the alphabet to represent tensor indices associated to the quotient bundle $T\scrI/n \to \scrI$. E.g, since $n^a h_{ab}=0$ then $h_{ab}$ defines an invertible conformal metric $\hConf_{AB} \in \So{S^2(T\scrI/n)^*\otimes L^{2}_{\scrI} }$. Finally, we will write $\gth^A_a$ the canonical projection $\gth^A_a \from T\scrI \to T\scrI/n$. E.g. $\hConf_{AB}\gth^A_a\gth^B_b = \hConf_{ab} $, $\bn^a \gth_a^A =0$.

\paragraph{Ehresmann connections}

An Ehresmann connection is a choice of horizontal distribution $H_x$ at each point $x$ of $\scrI$, i.e. such that $T_x\scrI = Span(n)_x \oplus H_x$. It amounts to a choice of embedding $m_A{}^{\mu} \from T_x\scrI/n \to T_x \scrI\subset T_xM$ such that the image has maximal rank and no intersection with the line generated by $n^a$. With these notations, we have:
\begin{equation*}
n^{\mu} = g^{\mu\nu} d_{\nu}\gO, \qquad n^{\mu}m_A{}_{\mu}=0,\qquad m_A{}^{\mu}m_B{}_{\mu} =h_{AB}.	
\end{equation*}
This uniquely defines a ``null tetrad'' $\left(l^{\mu}, m^{\mu}_A, n^{\mu} \right)$ at $\scrI$, by requiring that
\begin{align*}
l^{\mu}n_{\mu}=1,\qquad l^{\mu}l_{\mu}= l^{\mu}m_A{}_{\mu}= 0.
\end{align*}
Note that here ``null tetrads'' amount to isomorphisms $T\scrI \to \bbR \oplus T\scrI/n \oplus \bbR$ and are strictly less information than null tetrads in the more usual sense. 

One can typically obtain such null tetrads from a choice of local coordinates $\left(l^{\mu}, m^{\mu}_A, n^{\mu} \right) ``=\text{''} \left(\parD^{\mu}_{\gO} , \parD^{\mu}_A, \parD^{\mu}_u \right)$.  We will thus write suggestively any vector field $\Phi^{\mu}$ at $\scrI$ as $\Phi^{\mu} = \Phi^{\gO} l^{\mu} + \Phi^A m_A{}^{\mu}+ \Phi^u n^{\mu}$ and any 1-form $\Phi_{\mu}$ as $\Phi_{\mu} = \Phi_{\gO} n_{\mu} + \Phi_A m^A{}_{\mu}+ \Phi_u l_{\mu}$. We emphasise that this notation is only suggestive and that no choice of local coordinates is implied here. The only choices that are made are a choice of representative $g_{\mu\nu} \in \gConf_{\mu\nu}$ for the conformal metric and the choice of Ehresmann connection $m_A{}^{\mu}$.

Finally, in order to lighten future expressions we will sometimes use the following convention: we will write Lie derivatives along $n$ with a dot and we will use the short hand $\covD_{A} := \covD_{m_A}$. E.g. if $f\in \Co{\scrI}$ we will write its exterior derivative $(df)_a = \fd l_{a} + \covD_A f \gth^A_a$. We will also write $\covD_A$ for the ``horizontal'' covariant derivative $\covD^{(h,m)}_A$ induced on $T\scrI/n$ by the Levi-Civita connection of $h_{AB}$ and the choice of Ehresmann connection.

\subsubsection{Einstein's equations at lowest order}\label{sssection: Einstein's equations at lowest order}

Let us now consider what the asymptotic condition $D_{\gr}I\left(\gO\right)^I =O\left(\gO^2\right)$ implies on $\scrI$. Let $K_a{}^b := \covD_a n^b$ be the extrinsic curvature and  $\Kr_a{}^b :=  K_{a}{}^b -\tfrac{1}{d-1}\gd_a{}^b K_c{}^c$ its trace-free part. Let us also define the weighted scalar $\gk$ as the eigenvalue $n^a \Kr_{a}{}^b= \gk \; n^b$.  We leave this as an exercise to the reader to check that the extrinsic curvature is symmetric $K_{ab} = K_{ba}$, has the normal for eigen-vector and that $K_{ab} = \tfrac{1}{2}\LieD_{n} h_{ab}$, see also \cite{galloway_null_2004,gourgoulhon_31_2006} for more details and see appendix \ref{section: Appendix, BMS expansion and Einstein equations} for expressions in BMS coordinates.

  With a bit of work one can show from the definition of the normal connection \eqref{Asymptotically flat spacetimes: Tractor connection, Definition} that
\begin{align}\label{zeroth order structure: DI(omega)}
 D_{a}I(\gO)^I\big|_{\gO =0} &= \left. \Mtx{0 \\ \Kr{}_{a}{}^b - \tfrac{\gk}{d}\gd_a{}^b   \\[0.2cm] -\tfrac{1}{d-2}\covD_c \Kr_a{}^c - \tfrac{1}{d} \covD_a \gk} \right|_{\gO =0} 
\end{align}
 In order to obtain \eqref{zeroth order structure: DI(omega)}, it is useful to note that we have the identity $\tfrac{1}{d}\covD_{\mu} n^{\mu} = \tfrac{1}{d}\kappa + \tfrac{1}{d-1}K_a{}^a$. The last line of \eqref{zeroth order structure: DI(omega)} is obtained by making use of a null version of the Gauss-Codazzi equation: $\covD_{[a} K_{b]}{}^c = \tfrac{1}{2} R^c{}_{dab} n^d$.

The vanishing of \eqref{zeroth order structure: DI(omega)} is equivalent to the vanishing of the extrinsic trace-free curvature $\Kr_{a}{}^{b}$. In other terms, the infinity tractor is covariantly constant along $\scrI$ if and only if $\scrI$ is umbilic. Note that since we obtained this result by means of the tractor calculus, this is a direct proof that this is a conformally invariant property. Making a choice of Ehresmann connection $m_A{}^{\mu}$ we can write
\begin{align}\label{Zeroth order structure at null infinity: trace-free extrinsic curvature}
K_a{}^b &= h^{BC}\Big( \tfrac{1}{2} \hd_{AC} \Big) \; \gth^A_a \;  m_B{}^b + \Big(K_{A}{}^u\; \gth^A_a + \big(\tfrac{h^{CD}\hd_{CD}}{2(d-2)} +  \tfrac{d-1}{d-2}\gk\big) \; l_{a} \;\Big) n^b 	\\
\Kr_{a}{}^b &= h^{BC}\left( \tfrac{1}{2} \hd_{AC}\big|_0 - \tfrac{\gk}{d-2} h_{AC} \right) \; \gth^A_a \;  m_B{}^b + \Big(K_{A}{}^u\; \gth^A_a + \gk \; l_{a} \;\Big) n^b.
\end{align}
Here (and everywhere in this article) we use the notation $\hd_{AB} := \LieD_{n}h_{AB}$ and $\hd_{AB}\big|_0 := \hd_{AB} - \tfrac{h^{CD}\hd_{CD}}{d-2} h_{AB}$. On top of imposing $\gk =0$, Einstein's equations at lowest order thus constraint the metric $h_{AB}$ to be constantly dragged, up to an overall factor, along the null generators of $\scrI$.
\begin{Definition}
	We will say that a conformal Carroll geometry $\left(\scrI , \hConf_{ab} , \bn^a\right)$ is of null infinity type if $\LieD_{\bn}\hConf_{ab} \propto \hConf_{ab}$ (equivalently $\bhd_{AB}\big|_0=0$).
\end{Definition}
By the above discussion, conformal Carroll geometry of null infinity type are precisely those that can taken as ``seeds'' for asymptotically flat spacetimes. Consequently we will only consider such geometries in what follows. These are also such that $\gS$ is equipped with a conformal metric $\bh_{AB} \in \So{S^2 T^*\gS \otimes (L_{\gS})^2}$. This is because, for any Conformally Carroll Geometry $\left(\scrI = \bbR \times \gS , \hConf_{ab} , \bn^a\right)$, one can always choose $h_{ab} \in \bh_{ab}$ such that $h^{CD}\hd_{CD}=0$ (equivalently, assuming $\gk=0$, $\covD_{\mu}n^{\mu}\big|_{\scrI}=0$ ) and such a choice is unique up to
\begin{equation*}
 h_{AB} \mapsto (\go_{0})^2 h_{AB}
\end{equation*}
with $\go_{0} \in \So{L_{\gS}}$.

We close on a final remark about the geometrical meaning of $\bgk \in \So{L^{-1}}$, the eigen-value of the trace-free extrinsic curvature in the normal direction, $n^a \Kr_{a}{}^b= \gk \; n^b$. As discussed in \cite{bailey_thomass_1994} if $\gO$ is a defining function for an hyper-surface $\scrI \subset M$, its normal tractor $N^I$ is
\begin{equation*}
N^I := \Mtx{ 0 \\ n^a \\ -\tfrac{1}{d-1} K_a{}^a}.
\end{equation*}
It was shown in this reference that $\scrI$ is umbilic if and only if the normal tractor is parallel transported along $\scrI$. Making use of the identity $\tfrac{1}{d}\covD_{\mu} n^{\mu} = \tfrac{1}{d}\kappa + \tfrac{1}{d-1}K_a{}^a$, one obtains
\begin{equation*}
N^I - I\left(\gO\right)^I = \tfrac{1}{d} \gk X^I.
\end{equation*}
Therefore $\bgk$ parametrizes the discrepancy between the normal and infinity tractors. 

\subsection{The induced tractor bundle at null infinity}

\subsubsection{Null-tractors}

Following \cite{branson_conformally_2001,gover_almost_2010,curry_introduction_2018} it is tempting to identify the null-tractor bundle $\cT_{\scrI} \to \scrI$ with $I^{\perp} \to \scrI$ the orthogonal complement to the infinity tractor $I^I$. This is indeed a good guess. There are however crucial differences due to the fact that $\scrI$ is a null hypersurface. The most obvious one is that, since $I^I$ is null, the restriction of the tractor metric $g_{IJ}$ to $\cT_{\scrI}$ is degenerate with degenerate direction spanned by $I^I$. For this reason we will say that sections of $\cT_{\scrI}$ are ``null-tractors''. 

More subtle are the induced transformation rules under a conformal rescaling $g_{\mu\nu} \mapsto \go^{2} g_{\mu\nu}$: as we shall see shortly these do not only depend on the leading order term in the expansion $\go = \go_0 + \gO\; \go_1 + o\left(\gO\right)$, but also on the subleading order term $\go_1$. This fact makes the interpretation of the transformation rules for null-tractors less obvious. The freedom in $\go_0$ indeed is straightforwardly interpreted as freedom in rescaling $h_{ab}$, the degenerate metric at $\scrI$ but how are we to interpret the freedom in $\go_1$ ? 

A hint at the solution is given by considering the following: the freedom in $\go_0$ amounts to freedom in trivialisations of $L_{\scrI} \to \scrI$. However $\scrI$ is itself (at least locally) a trivial fibre bundle $\scrI \to \gS$ and we will see that the freedom in $\go_1$ can be parametrized as a freedom in choosing a trivialisation $u_0 \from \scrI \to \bbR$ for this bundle. For any $u_0 \in  \Co{\scrI}$ such that $\go_0 := \covD_{n}u_0$ is nowhere vanishing, we will in fact construct a map 
\begin{align*}
\begin{array}{ccccc}
 \Co{\scrI} & \to &  \So{L} \\
u_0 & \mapsto & \go\left(u_0\right) = \go_0 + \go_1 +o(\gO).
\end{array}
\end{align*}
Crucially, even thought the construction will make use of the interior metric $\left(M,\gConf_{\mu\nu}, \bgO\right)$, the resulting $\go_1$ will not depend on the details of $\gConf_{\mu\nu}$. This will ensure that the resulting construction is intrinsic to $\scrI$.

As we already pointed out, details on the intrinsic construction of null-tractors solely in terms of the conformal Carroll geometries (i.e. without the need to refer to the spacetime geometry) were discussed in \cite{herfray_asymptotic_2020}. These were however presented in the gauge where $\tfrac{1}{d-2}h^{CD}\hd_{CD}$ is null while this will be relaxed in this presentation. We will however show that the two constructions match each others under the condition $h^{CD}\hd_{CD}=0$ (equivalently $\covD_{\mu}n^{\mu}\big|_{\scrI}=0$).

\subsubsection{Splitting isomorphism and transformation rules}\label{sssection: Transformation rules}

Let $m_A{}^{\mu}$ be a choice of Ehresmann connection and $\left(l^{\mu}, m_A{}^{\mu}, n^{\mu}\right)$ the associated null tetrad (see the discussion in section \ref{sssection: The quotient tangent bundle and Ehresmann connections}). Recall that if $\Phi^{\mu}$ is a section of the tangent bundle at $\scrI$, we write  $\Phi^{\mu} = \Phi^{\gO} l^{\mu} + \Phi^A m_A{}^{\mu}+ \Phi^u n^{\mu}$. Accordingly, a generic tractor section $\Phi^I$ at $\scrI$ will be written as 
\begin{equation*}
\Phi^I \;\xeq{g,m}\; \Mtx{ \Phi^+ \\ \Phi^{\gO} \\ \Phi^A \\ \Phi^u \\ \Phi^- } \quad\text{with}\quad  g_{IJ} \;\xeq{g,m}\; \Mtx{0 & 0 & 0 & 0 & 1 \\ 0 &0 & 0 & 1 &0 \\ 0& 0& h_{AB} & 0 & 0 \\ 0 & 1 & 0&0 &0 \\
1&0&0&0&0}, \qquad I^I \;\xeq{g,m}\; \Mtx{ 0 \\ 0 \\ 0 \\ 1 \\ -\tfrac{1}{2(d-2)}h^{CD}\hd_{CD}}.
\end{equation*}
 It follows that $\Phi^I$ belongs to $I^{\perp}$ if and only if $\Phi^{\gO} + \Phi^+ I^- =0$. 

We emphasise that the only choices that are made here are a choice of representative $g_{\mu\nu} \in \gConf_{\mu\nu}$ for the conformal metric and the choice of Ehresmann connection $m_A{}^{\mu}$.

\paragraph{Null-tractors} If $\Phi^I$ is a section of $I^{\perp}$ we define the associated null-tractor $\Phit^I \in \So{\cT_{\scrI}}$ through the isomorphism
\begin{align}\label{Zeroth order structure at null infinity: Null-Tractors, isomorphism}
	\Phi^I \;\xeq{g,m}\; \Mtx{\Phit^+ \\ - I^- \Phit^+ \\ \Phit^A \\ \Phit^u \\ \Phit^- + I^- \Phit^u} \in I^{\perp} \qquad \simeq \qquad \Phit^I \;\xeq{g,m}\;\Mtx{\Phit^+ \\ \Phit^A \\ \Phit^u \\ \Phit^-} \in \cT_{\scrI}.
\end{align}
In other terms, while a choice of representative $g_{\mu\nu} \in \gConf_{\mu\nu}$ gave an isomorphism $\cT \stackrel{g}{=} \bbR \oplus TM \oplus \bbR$ for the tractor bundle $\cT \to M$, here a choice of pair $\left(g_{\mu\nu}, m_A{}^{\mu}\right)$ gives an isomorphism $\cT_{\scrI} \stackrel{g, m}{=} \bbR \oplus T\scrI/n \oplus \bbR \oplus \bbR$  for null-tractors $\cT_{\scrI} \to \scrI$. This isomorphism is such that  
\begin{equation}\label{Zeroth order structure at null infinity:  Null tractors, metric , X and I}
	h_{IJ} \;\xeq{g,m}\; \Mtx{0 &  0 & 0 & 1 \\ 0& h_{AB} & 0 & 0 \\ 0 & 0&0 &0 \\
		1&0&0&0}, \qquad \It^I \;\xeq{g,m}\; \Mtx{ 0 \\ 0  \\ 1 \\0}, \qquad \Xt^I \;\xeq{g,m} \; \Mtx{0 \\ 0\\ 0\\ 1}
\end{equation}
where $h_{IJ}$ is the induced metric on null-tractors. It is degenerate with kernel spanned by $\It^I$. 

\paragraph{Dual null-tractors} Dual null-tractor $\left(\cT_{\scrI}\right)^*$ are canonically isomorphic to the quotient bundle $\cT / I$. As a convention we will write,
\begin{align*}
	\Psi_I \;\xeq{g,m}\; \sMtx{\Psit_- \\ \Psit_u - I^{-} \Psit_{-} \\ \Psit_A \\ \Psit_{\gO}\\ \Psit_{+} + I^- \Psit_{\gO} }\in \cT/I \qquad \simeq \qquad \Psit_I \;\xeq{g,m}\;\Mtx{\Psit_{-} \\ \Psit_u  \\ \Psit_A \\ \Psit_{+}} \in (\cT_{\scrI})^*.
\end{align*}
with pairing 
\begin{equation*}
\Phit^I\Psit_I = \Phit^+\Psit_{+} + \Phit^A\Psit_A +\Phit^u \Psit_u + \Psit^-\Phit_{-}.
\end{equation*}

\paragraph{Transformation rules} 
As we already pointed out, the splitting isomorphism $\cT_{\scrI} \stackrel{g, m}{=} \bbR \oplus T\scrI/n \oplus \bbR \oplus \bbR$ of null-tractors relies on both the choice of $g_{\mu\nu} \in \gConf_{\mu\nu}$ and $m_A{}^{\mu}$. In other terms, the transformation rules
\begin{equation*}
g_{\mu\nu} \mapsto \ggh_{\mu\nu} = \go^2 g_{\mu\nu},\qquad \gO \mapsto \gOh = \go \gO, 	
\end{equation*}
must be supplemented by the change of Ehresmann connection $\scrI$
\begin{equation}\label{Zeroth order structure at null infinity: transformation rules- ehresmann connection}
\left(n^{\mu} , m_A{}^{\mu}\right) \mapsto \left(\nh^{\mu} , \mh_A{}^{\mu}\right) =\left(\go_0^{-1} n^{\mu} \;,\; m_A{}^{\mu} - t_A n^{\mu} \right).	
\end{equation}
Here $\go = \go_0 + \gO \go_1 +o(\gO)$ is a nowhere vanishing function on $M$ and $\go_0$, $\go_1$, $t_{A}$ are respectively functions and a tensor on $\scrI$. The transformation rules \eqref{Asymptotically flat spacetimes: Tractors, transformation rules} for tractors then give the following transformation rules for boundary tractors
\begin{align*}
	\Phi^I\big|_{\scrI} &\stackrel{g,\;m}{=}\; \Mtx{\Phi^+ \\ \Phi^{\gO} \\ \Phi^{A} \\ \Phi^u \\ \Phi^-} &\mapsto&&  \Phih^I\big|_{\scrI} &\stackrel{\ggh,\;\mh}{=}\;
	\Mtx{ \go_0 & 0 & 0& 0 & 0 \\[0.3em]
		\tfrac{\god_0}{\go_0} &1 &  0& 0 & 0\\[0.3em]
		- \tfrac{\god_0}{\go_0} \tfrac{t^A}{\go_0} +\tfrac{\gU^A }{\go_0}& -\frac{t^A}{\go_0} &	\tfrac{1}{\go_0} \gd^{A}{}_{B} &0 & 0\\[0.3em]
- \tfrac{1}{2} t^C t_C\tfrac{\god_0}{\go_0} + t_C \gU^C + \tfrac{\go_1}{\go_0}	& - \tfrac{1}{2}t^C t_C &	t_B &1& 0	\\[0.3em]
		-\tfrac{\god_0}{\go_0} \tfrac{\go_1}{(\go_0)^2} - \tfrac{\gU^A \gU_A}{\go_0}& -\tfrac{\go_1}{(\go_0)^2}& -\tfrac{\gU_B}{\go_0}& -\tfrac{\god_0}{(\go_0)^2} & \go^{-1}_0}
\Mtx{\Phi^+ \\ \Phi^{\gO} \\ \Phi^{B} \\ \Phi^u \\ \Phi^-}
\end{align*}
Where we introduced the notation $\gU_A := \go_0^{-1} \covD_{A} \go_0$ and $\tfrac{\god_0}{\go_0} := \go_0^{-1} \covD_{n} \go_0$.

 Making use of the isomorphism \eqref{Zeroth order structure at null infinity: Null-Tractors, isomorphism}, we obtain the transformation rules for null-tractors
\begin{align}\label{Zeroth order structure at null infinity: Null-Tractors, transformation rules}
\Phit^I &\stackrel{g,\;m}{=}\; \Mtx{\Phit^+ \\ \Phit^{A} \\ \Phit^u \\ \Phit^-} &\mapsto& & \Phith^I &\stackrel{\ggh,\;\mh}{=}\;
\Mtx{ \go_0 & 0 & 0& 0 \\
	\go_0^{-1} U^{A} & 	\go_0^{-1} \gd^{A}{}_{B} &0 & 0\\
	\gb& 	t_B &1& 0	\\
	-\go^{-1}_0\;\tfrac{1}{2} U^C U_C & -\go^{-1}_0\;U_{B}&0& \go^{-1}_0}
\Mtx{\Phit^+ \\ \Phit^{B} \\ \Phit^u \\ \Phi^-}
\end{align}
where $\qquad U_A := \gU_A +  \left(I^- - \tfrac{\god_0}{\go_{0}}\right) t_A \quad $ and $ \quad \gb := \tfrac{\go_1}{\go_0} + t^C \gU_C +\tfrac{1}{2} t^C t_C \left( I^- - \tfrac{\god_0}{\go_0} \right)$. We recall from the previous section that $I^- = -\tfrac{1}{2(d-2)}h^{CD}\hd_{CD}$.

 The dependence of the isomorphism \eqref{Zeroth order structure at null infinity: Null-Tractors, isomorphism} on a choice of Ehresmann connection, even thought qualitatively different from usual tractors, is intrinsic to the boundary. The appearance of $\go_1$ in the transformation rules for null-tractors is more problematic for it prevents us to interpret these as geometrical objects intrinsic to the boundary. 

We will now show that both choices can be canonically parametrized by a choice of trivialisation $u\in \Co{\scrI}$.

\subsection{\texorpdfstring{Trivialisations of $\scrI$ and BMS coordinates}{Trivialisations of scri and BMS coordinates}}

\subsubsection{\texorpdfstring{Trivialisations of $\scrI$}{Trivialisations of scri}}

Recall that we suppose that $\scrI$ is the total space of a trivial fibre bundle $\scrI \to \gS$ whose null fibres are generated by $n^a$ (this is however purely for convenience since all results are local in nature).

\begin{Definition}
A choice of \emph{trivialisation} for a conformal Carroll geometry $\left(\scrI , \bh_{ab} , \bn^a\right)$ is a choice of function $u \in \Co{\scrI}$ such that $\bgs^{-1} := \covD_{\bn} u$ is nowhere vanishing.

 In particular a trivialisation defines preferred representatives $\left(h_{ab},n^a\right):=\left( \bgs^{-2}\hConf_{ab} ,  \bgs\bn^a\right)$. The resulting compatible triplet $\left( u , h_{ab}, n^a\right)$ was called well-adapted trivialisation of conformal Carroll geometry in \cite{herfray_asymptotic_2020}.
\end{Definition}
\begin{Lemma}\label{Zeroth order structure at null infinity: Lemma - trivialisation transformation rules}
	Let $u$ and $\uh:=u_0$ be two trivialisations of $\left( \scrI, \bh_{ab}, \bn^a\right)$ and let $\Big(h_{ab} ,n^a\Big)$, $\left(\hh_{ab},\nh^a\right)$ be the associated representatives. The transformation rules can be written as follows
	\begin{align}\label{Zeroth order structure at null infinity: trivialisation, transformation rules}
		\Big(u, h_{ab}, n^a\Big)&& \mapsto&& \Big(\uh, \hh_{ab}, \nh^a \Big) =&\Big(u_0, (\go_0)^2 h_{ab}, (\go_0)^{-1} n^a \Big).
	\end{align}
	where $\go_0 := \covD_{n} u_0 $.
\end{Lemma}
It was shown in \cite{herfray_asymptotic_2020} that choices of well-adapted trivialisation give isomorphisms and transformation rules for null-tractors i.e.
\begin{align*}
\Phit^I &\stackrel{u}{=}\; \Mtx{\Phit^+ \\ \Phit^{A} \\ \Phit^u \\ \Phit^-} &\mapsto& & \Phith^I &\stackrel{\uh}{=}\;
M\left( u_0 \right)
\Mtx{\Phit^+ \\ \Phit^{B} \\ \Phit^u \\ \Phit^-}.
\end{align*}
Where $M\left( u_0 \right)$ is a matrix function of $u_0$. We will here re-establish this result by making use of the transformation rules for null-tractors \eqref{Zeroth order structure at null infinity: Null-Tractors, transformation rules} which where obtained from the spacetime geometry. Note that, as opposed to \cite{herfray_asymptotic_2020}, we here do not require trivialisations to satisfy $h^{CD}\hd_{CD}= 0$.

Clearly, any trivialisation $u \in \Co{\scrI}$ defines an horizontal distribution $H_x := Ker\left(du\right)_x$ and therefore an Ehresmann connection $m_A{}^{\mu}$. Considering $u$ and $\uh$ as in lemma \ref{Zeroth order structure at null infinity: Lemma - trivialisation transformation rules}, we have:
\begin{equation*}
\left(n^{\mu} ,m_A{}^{\mu}\right) \mapsto \left(\nh^{\mu} , \mh_A{}^{\mu}\right) =\left(\go_0^{-1} n^{\mu} \;,\; m_A{}^{\mu} - \go_0^{-1}\covD_{A} u_0\; n^{\mu} \right).	
\end{equation*}
Matching these with the similar expression \eqref{Zeroth order structure at null infinity: transformation rules- ehresmann connection} in the previous section, we make the identification $t_A = \tfrac{\covD_{A} u_0}{\go_0}$. We will now show that the term $\go_1$ in the transformation rules \eqref{Zeroth order structure at null infinity: Null-Tractors, transformation rules} for null-tractors can also be geometrically parametrised by $u_0$. As a consequence, the transformation rules will be parametrised in terms of data at $\scrI$ only, turning null-tractors into intrinsic geometrical objects. 

\subsubsection{BMS coordinate and scale induced by a choice of trivialisation}

Let $u \in \Co{\scrI}$ be a trivialisation of $\scrI$. We wish to prove that it uniquely defines a representative $g_{\mu\nu} \in \gConf_{\mu\nu}$ in a neighbourhood of $\scrI$. In other words that a trivialisation picks, in a neighbourhood of $\scrI$, a preferred scale for the conformal metric.  It will follow that,  in a neighbourhood of $\scrI$, a change of trivialisations \eqref{Zeroth order structure at null infinity: trivialisation, transformation rules} will uniquely parametrize a change of scale $g_{\mu\nu} \mapsto \go^2 g_{\mu\nu} $ (and therefore uniquely parametrize tractor transformation rules \eqref{Asymptotically flat spacetimes: Tractors, transformation rules}). In particular, this will provide the explicit parametrization
\begin{equation*}
\go\left(u_0\right) = \go_0\left(u_0\right) + \go_1\left(u_0\right) \gO + o\left(\gO\right).
\end{equation*}
Evaluating the result on \eqref{Zeroth order structure at null infinity: Null-Tractors, transformation rules} will give the transformation rules for null-tractors induced by change of trivialisations.

 The reader which is not interested in the derivation of these transformation rules can therefore skip this part and move directly to our summary of the results (sub-section \ref{ssection: Null tractors: summary}).

\paragraph{BMS coordinates} 
Choices of well-adapted trivialisations $u \in \Co{\scrI}$ at $\scrI = \bbR \times \gS$ are essentially equivalent to choices of BMS coordinates $\left(\gO, u ,\gth\right)$ in a neighbourhood $U$ of $\scrI$ - here $\gth$ is a map $\gth \from U \to \gS$. We here recall this classical construction, see \cite{tafel_comparison_2000,madler_bondi-sachs_2016} for modern discussions. 

Let $\left(\scrI, \bh_{AB} , \bn^a \right)$ be a conformal Carroll structure of null infinity type and let $\left(\MConf, \gConf_{\mu\nu} , \bgO\right)$ be an asymptotically flat spacetime extending it. Let $u\in\Co{\scrI}$ be a trivialisation of $\scrI$ and $\left(n^a, h_{ab}\right)$ the corresponding representatives. Let $\left( \gO , g_{\mu\nu} \right)$ be choices of representatives such that $\gi^* g_{\mu\nu}= h_{ab}$, $n^{\mu} = g^{\mu\nu}(d\gO)_{\nu}\big|_{\scrI} $. These are unique up to $\left( \gO, g_{\mu\nu}\right) \mapsto \left(\go \gO, \go^2 g_{\mu\nu} \right)$ with $\go\big|_{\scrI} =1$. There is a unique vector field $l^{\mu}$ at $\scrI$ such that $l^{\mu}l_{\mu}=0$, $l^{\mu} n_{\mu}=1$ and $l^{\mu}(du)_{\mu}=0$. This vector field is pointing ``outside'' of the null boundary and one can therefore consider the set of null geodesics that it generates. Since null geodesics are conformally invariant this construction does not depend on the choice of representatives $\left(\gO, g_{\mu\nu}\right)$. In a suitable neighbourhood $U$ of $\scrI$, each point lies on a unique null geodesic. Since the set of null geodesics are parametrised by $\scrI$ -which is itself identified with $\bbR \times \gS$ by the choice of trivialisation $u$- this defines a map $\gth \from U \to \gS$. Together with the boundary defining function $\gO \in \Co{M}$, this yields a set of Gaussian null coordinates $\left(\gO, u, \gth \right)$ and one can write
\begin{equation}\label{Zeroth order structure at null infinity: Well-adapted trivialisation, BMS coordinates}
g_{\mu\nu} = \gO^3 e^{2\gb}  V (du)_{\mu}(du)_{\nu} + e^{2\gb} 2 (du)_{\mu} (d\gO)_{\nu} + H_{AB}\left((d\gth)_{\mu}{}^{A} -U^A (du)_{\mu} \right)\left( (d\gth)_{\nu}{}^{B} -U^B (du)_{\nu} \right)
\end{equation}
where $\beta$, $\gO^3 V$, $U^A$ and $H_{AB}$ are functions on $U \subset M$. By construction, one has $\gb \big|_{\scrI}=0$, $\gO^3 V\big|_{\scrI}=0$, $U^A\big|_{\scrI}=0$ and $H_{AB}\big|_{\scrI} = h_{AB}$. At this stage the coordinate system is fixed uniquely up to the remaining ambiguity $\left( g_{\mu\nu} , \gO\right) \mapsto \left( \go^2 g_{\mu\nu} , \go \gO\right)$ with $\go\big|_{\scrI} =1$. One fixes this ambiguity by requiring $\parD_{\gO}det(H_{AB}) =  0$. We therefore obtained the following:
\begin{Proposition}\label{Zeroth order structure at null infinity: Well-adapted trivialisation, BMS coordinates Proposition}
	Trivialisations $u\in\Co{\scrI}$ on the boundary of an asymptotically flat spacetimes $\left(\MConf, \bgg_{\mu\nu} ,\bgO \right)$ are in one to one correspondence with choices of BMS coordinates i.e. local coordinates $\left(\gO , u, \gth\right)$ in a neighbourhood of $\scrI$ such that $g_{\mu\nu}$ can be written as \eqref{Zeroth order structure at null infinity: Well-adapted trivialisation, BMS coordinates} with $\parD_{\gO}det(H_{AB}) =  0$.
\end{Proposition}

Since one of the BMS coordinates is a representative $\gO \in \bgO$, the above proposition implies that well-adapted trivialisation effectively pick a scale in a neighbourhood of $\scrI$. We will now consider the induced transformation rules.

\paragraph{Change of scale induced by a change of well-adapted trivialisation}

Let $u \mapsto \uh := u_0 $ be a change of well-adapted trivialisation and $\go_0 := \covD_{n}u_0$. By proposition \ref{Zeroth order structure at null infinity: Well-adapted trivialisation, BMS coordinates Proposition} this corresponds to a change of BMS coordinates $\left(\gO , u , \gth \right) \mapsto \left(\gOh = \go\gO  , \uh , \gthh = f\left(\gth\right) \right)$ -where $f \from \gS \to \gS$. To leading order we have
\begin{align}\label{Zeroth order structure at null infinity: Well-adapted trivialisation, Change of BMS coordinates, expansion}
\goh &= \go_0 + \gO \;\go_1\left(u_0\right)  + o\left(\gO\right), \\
\uh &= u_0 + \gO \;u_1\left(u_0\right) + o\left(\gO\right),\nonumber\\
f&= Id_{\gS}+ \gO \;f_1^A \left(u_0\right) + o\left(\gO\right).\nonumber
\end{align}

\begin{Proposition}\label{Zeroth order structure at null infinity: Proposition Well-adapted trivialisation, Main Proposition}
	Let $\left(\scrI,\hConf_{ab},\bn^a\right)$ be a conformally Carroll structure of null infinity type on a $(d-1)$-dimensional manifold $\scrI$. Let $\left(\MConf, \gConf_{\mu\nu}, \bgO\right)$ be an asymptotically flat spacetime (such that $\gO^{-2} D_{\gr}I\left(\gO\right)^I$ has a finite smooth limit at $\scrI$) extending $\left(\scrI,\hConf_{ab},\bn^a\right)$.
	
	Let $u$ and $\uh:=u_0$ be two well-adapted trivialisations for $\scrI$. Let their respective representatives $\big(n^a, h_{ab}\big)$, $\big(\nh^a, \hh_{ab}\big)$ be related by \eqref{Zeroth order structure at null infinity: trivialisation, transformation rules}. Let $\big(\gO, u, \gth \big)$ and $\big(\gOh, \uh, \gthh\big)$ be the respective BMS coordinates given by proposition \ref{Zeroth order structure at null infinity: Well-adapted trivialisation, BMS coordinates Proposition}. Then these two coordinates systems are asymptotically related by \eqref{Zeroth order structure at null infinity: Well-adapted trivialisation, Change of BMS coordinates, expansion} with
	\begin{align*}
	f^A_1 &= -\tfrac{1}{\go_0} \covD^A u_0, & u_1 &= -\tfrac{1}{2\go_0} \covD^C u_0 \covD_C u_0,
	\end{align*}
	and
	\begin{equation*}
	\go_1 = -\tfrac{1}{d-2} \covD^C \covD_C u_0 
	- \tfrac{d-4}{d-2} \tfrac{1}{\go_0} \covD^C u_0 \covD_C \go_0  + \tfrac{d-4}{d-2} \tfrac{1}{2 \go_0} (\covD u_0)^2 \left( \tfrac{\god_0}{\go_0} + \tfrac{1}{2(d-2)}h^{CD}\hd_{CD} \right).
	\end{equation*}

In particular the leading terms in the asymptotic expansion \eqref{Zeroth order structure at null infinity: Well-adapted trivialisation, Change of BMS coordinates, expansion} do not depend on the choice of $\left(\MConf, \gConf_{\mu\nu} , \bgO\right)$ extending $\left( \scrI, \hConf_{ab}  ,\bn^a\right)$.

\end{Proposition}

The proof of this proposition is postponed in appendix \ref{section: Appendix, Proof of BMS-transformation Proposition}.

\subsection{Null tractors: summary}\label{ssection: Null tractors: summary}

\subsubsection{Definition and transformation rules}
Let us here summarize what has been achieved in this section.

 Let $\left(\scrI,\hConf_{ab},\bn^a\right)$ be a conformal Carroll geometry of null infinity type and let $\left(\MConf, \gConf_{\mu\nu}, \bgO\right)$ be an asymptotically flat manifolds extending it with $D_{\gr}I\left(\gO\right)^I = O\left(\gO^2\right)$. We defined the null-tractor bundle $\cT_{\scrI} \to \scrI$ has as the sub-bundle $I^{\perp}$ of the restriction  $\cT\big|_{\scrI}$ of the tractor bundle of $M$ to $\scrI$. We then showed that any choice of trivialisation, defined as a function $u\in \Co{\scrI}$ such that $\bgs^{-1}:=\covD_{\bn}u$ is nowhere vanishing, gives representatives $\left(h_{ab} , n^a\right) := \left(\bgs^{-2}\bh_{ab} , \bgs \bn^a \right)$ and a splitting isomorphism $\cT_{\scrI} \xeq{u}\bbR \oplus T\scrI/n \oplus \bbR \oplus \bbR$. Practically, if $\Phit^I \in \So{\cT_{\scrI}}$ is a section of the null-tractor bundle, we write
\begin{equation*}
\Phit^I \xeq{u} \Mtx{\Phit^+ \\ \Phit^A \\ \Phit^u \\ \Phit^-}.
\end{equation*}
The null-tractor bundle is equipped with a degenerate metric,
\begin{equation*}
\Phit^I \Phit^J h_{IJ} := 2\Phit^+\Phit^- + \Phit^A\Phit^B h_{AB}
\end{equation*}
and two preferred sections $\Xt^I\in \So{\cT_{\scrI} \otimes L_{\scrI} }$, $\It^I \in \So{\cT_{\scrI}}$ given by \eqref{Zeroth order structure at null infinity:  Null tractors, metric , X and I}.

If $\uh = u_0$ is any other trivialisation, the associated representatives are related via $\left(\hh_{ab},\nh^a\right) = \left((\go_0)^{2} h_{ab} , (\go_0)^{-1} n^a\right)$ with $\go_0:=\ud_0 \; (=\!\!\covD_{n}u_0)$. Finally, the isomorphism $\cT_{\scrI} \xeq{\uh}\bbR \oplus T\scrI/n \oplus \bbR \oplus \bbR$ is related to the initial one via transformation rules obtained by evaluating equation \eqref{Zeroth order structure at null infinity: Null-Tractors, transformation rules} for $t_A = \tfrac{\covD_{A}u_0}{\go_0}$ and for $\go_1$ given by proposition \ref{Zeroth order structure at null infinity: Proposition Well-adapted trivialisation, Main Proposition}. We here gather the end result:
\begin{align}\label{Zeroth order structure at null infinity: Null-Tractors, full transformation rules}
	\Phit^I &\stackrel{u}{=}\; \Mtx{\Phit^+ \\ \Phit^{A} \\ \Phit^u \\ \Phit^-} &\mapsto& & \Phith^I &\stackrel{\uh}{=}\;
	\Mtx{ \go_0 & 0 & 0& 0 \\
		\go_0^{-1} U^{A} & 	\go_0^{-1} \gd^{A}{}_{B} &0 & 0\\
		\gb& \go^{-1}_0 \covD_{B}u_0&1& 0	\\
		-\go^{-1}_0\;\tfrac{1}{2} U^C U_C & -\go^{-1}_0\;U_{B}&0& \go^{-1}_0}
	\Mtx{\Phit^+ \\ \Phit^{B} \\ \Phit^u \\ \Phi^-}
\end{align}
\begin{align*}
	 U_A &:= \gU_A - \left(\tfrac{\god_0}{\go_{0}} + \tfrac{\gTh}{2}\right) \tfrac{\covD_{A}u_0}{\go_0}, & \gb &:= \tfrac{1}{d-2}\left[  -\tfrac{1}{\go_0} \covD^C \covD_C u_0 + 2 \gU^C \tfrac{\covD_{C}u_0}{\go_0}- \left(\tfrac{\covD u_0}{\go_0}\right)^2 \left( \tfrac{\god_0}{\go_0} + \tfrac{\gTh}{2}\right)\right] ,
\end{align*}
 with $\gU_A = \go^{-1}_0 \nabla_A \go_0$, $\tfrac{\god_0}{\go_0} := \go_0^{-1} \covD_{n} \go_0$ and $\gTh := \tfrac{1}{(d-2)}h^{CD}\hd_{CD}$.

Neither the transformation rules nor the definition of the tractor bundle depends on the choice of asymptotically flat spacetime $\left(\MConf, \gConf_{\mu\nu} , \bgO\right)$ extending $\left(\scrI,\hConf_{ab}  ,\bn^a \right)$. In fact we have more: The isomorphisms $\cT_{\scrI} \xeq{\uh}\bbR \oplus T\scrI/n \oplus \bbR \oplus \bbR$ do not depend on the detail of the extension $\left(\MConf, \gConf_{\mu\nu}, \bgO\right)$ either. This is perhaps not fully clear at this stage but will be clarified when we discuss Thomas operator. Altogether this shows that the null-tractor bundle is intrinsic to the conformal Carroll geometry  $\left(\scrI,\hConf_{ab}, \bn^a\right)$.

\paragraph{BMS symmetries}
If one restricts oneself to trivialisations $u$, $\uh := u_0$ such that the resulting representatives $h_{AB}$, $\hh_{AB} = (\go_0)^2 h_{AB}$ satisfy $h^{CD}\hd_{CD}=0$, $\hh^{CD}\hhd_{CD}=0$ then $0=\god_0 =\udd_0$. In other terms $u_0 = \go_0\left(u-\xi\right)$ where $\go_0$ and $\xi$ are functions on $\gS$: $\xi$ are the usual supertranslation and $\go_0$ Weyl rescalings. Since the whole formalism is covariant under diffeomorphisms of $\gS$ the action of superrotations is straightforward. In particular the action on the BMS group on the asymptotic shear \cite{barnich_finite_2016} will be recovered by applying the tractor transformation rules \eqref{Zeroth order structure at null infinity: Null-Tractors, full transformation rules} on the induced tractor connection.

When restricting to these gauges, one sees from the transformation rules \eqref{Zeroth order structure at null infinity: Null-Tractors, full transformation rules} that $\cT_{\scrI}/\It$ is canonically isomorphic to the pull-back bundle $\pi^* \cT_{\gS}$ of (usual) tractors on $\left(\gS,\bh_{AB}\right)$. One can also check that the transformation rules \eqref{Zeroth order structure at null infinity: Null-Tractors, full transformation rules} coincide with those of \cite{herfray_asymptotic_2020}.

\subsubsection{Thomas operator and invariant definition of null-tractors}\label{sssection: Thomas operator and invariant definition of null-tractors}

Results presented in this subsection are independent of the rest of the article and can be safely skipped by the reader interested by details on the induced tractor connection.

We first show that the isomorphisms $\cT_{\scrI} \xeq{u} \bbR \oplus T\scrI/n \oplus \bbR \oplus \bbR$ are fully intrinsic to the conformal Carroll geometry $\left(\scrI, \bh_{ab} ,\bn^a \right)$, i.e. do not depend on the choice of asymptotically flat spacetime extending it.

We then recall the invariant definition of null-tractors as a sub-bundle of the second jet bundle $J^2 L_{\scrI}$ and show that it can be canonically related to the spacetime tractor bundle, the later being defined in terms of the jet bundle $J^2 L\big|_{\scrI}$. This will essentially amount to the construction of an intrinsic Thomas operator for null-tractors in terms of the spacetime Thomas operator.

\paragraph{Spacetime Thomas operator}
Let $\bgs_0 \in \So{L_{\scrI}}$ be a section of the boundary scale bundle $L_{\scrI}$ and let $\gs = \gs_0 + \gO \gs_1 + o\left(\gO\right) \in \So{L}$ be an extension in $M$. Let $I\left(\gs\right)^I$ be the associated infinity tractor given by Thomas operator \eqref{Asymptotically flat spacetimes: Thomas operator, Definition}. We have,
\begin{Lemma}\label{Zeroth order structure at null infinity: spacetime Thomas operator}
	\begin{equation*}
		I\left(\gs\right)_I\big|_{\scrI} = \Mtx{\gs_0 \\ \gsd_0 \\ \covD_A\gs_0 \\ \gs_1 \\ -\tfrac{1}{d}\left(\covD^C \covD_C \gs_0 + 2\gsd_1 -\tfrac{d-4}{2}V_2 \gs_0 \right) -\tfrac{1}{2(d-2)}h^{CD}\hd_{CD}\;  \gs_1}.
	\end{equation*}
where $V_2 := -\tfrac{2}{d-2}P^h$ if $d\geq4$ and $V_2 := M$ if $d=3$.

\noindent (Here and everywhere below, $P^h= \tfrac{1}{d-3}R^h$ is the trace of the Schouten tensor on $\gS$ and $M$ is the 3D mass aspect.)
\end{Lemma}
\begin{proof}This can be obtained by a direct computation in coordinate by making use of the explicit results of appendix \ref{section: Appendix, BMS expansion of geometrical quantities}
\end{proof}

In particular, if $u_0\in \Co{\scrI}$ is a trivialisation of $\scrI$ and we take $\gs\left(u_0\right) = \go_0 + \gO \go_1 + o\left(\gO\right)$ where $\go_0 := \ud_0$ and $ \go_1\left(u_0\right)$ is given by proposition \ref{Zeroth order structure at null infinity: Proposition Well-adapted trivialisation, Main Proposition} we obtain a section $I\left(\gs\left(u_0\right)\right)^I\big|_{\scrI} \in \So{\cT\big|_{\scrI}}$ such that $I\left(\gs\left(u_0\right)\right)^+\big|_{\scrI} = \gs_0$ is nowhere vanishing on $\scrI$. Such a section amounts to the isomorphism $\cT\big|_{\scrI} \xeq{\bgs(u_0)} \bbR \oplus TM \big|_{\scrI} \oplus \bbR$ discussed in section \ref{sssection: The tractor bundle} (this is because, together with $X^I$, it defines a null frame, see e.g. \cite{herfray_einstein_2022} for a detailed discussion). Since, for $d\geq4$, none of this depends on the choice of asymptotically flat spacetime $\left(\MConf, \gConf_{\mu\nu} , \bgO\right)$ extending $\left(\scrI , \bh_{ab} ,\bn^a \right)$ it follows that, as previously claimed, the isomorphism $\cT_{\scrI} \xeq{u_0} \bbR  \oplus T\scrI /n \oplus \bbR \oplus \bbR$ is intrinsic to the conformal Carroll geometry. For $d=3$ the freedom in the isomorphism is parametrized by the 3D mass aspect.

\paragraph{Thomas operator for the null-tractor bundle}
Let $\bgs_0 \in \So{L_{\scrI}}$ be a section of the boundary scale bundle $L_{\scrI}$ and let $\gs = \gs_0 + \gO \gs_1 + o\left(\gO\right) \in \So{L}$ be an extension in $M$.
We identically have
\begin{align}
D_{\gr}I\left(\gs\right)^+ &=0, & D_{\mu}I\left(\gs\right)^{\mu} &=0.
\end{align}
When restricted to $\scrI$ the equation $n^{\gr} D_{\gr}I\left(\gs\right)_{\mu}\big|_{\scrI}=0$ is therefore conformally invariant and we have
\begin{Lemma}\label{Zeroth order structure at null infinity: intrisic operator, nDI}
	$\gs = \gs_0 + \gO \gs_1 + o\left(\gO\right) \in \So{L}$ is a solution of $n^{\gr} D_{\gr}I\left(\gs\right)_{a}\big|_{\scrI}=0$\\ if and only if
	\begin{align*}
\covD_a\left(\gsd_0 - \tfrac{1}{2(d-2)}h^{CD}\hd_{CD}\;\gs_0\right)&=0
	\end{align*}
and a solution of $n^{\gr} D_{\gr}I\left(\gs\right)_{\mu}\big|_{\scrI}=0$ if and only if, on top of the above equation,
\begin{align*}
\gsd_1& = \tfrac{1}{d-2}\covD^C\covD_C \gs_0 -V_2 \gs_0
\end{align*}
where $V_2 := -\tfrac{2}{d-2}P^h$ if $d\geq4$ and $V_2 := M$ if $d=3$.
\end{Lemma}
\begin{proof}This is again obtained by a direct computation in coordinate and making use of the explicit results of appendix \ref{section: Appendix, BMS expansion of geometrical quantities}.
\end{proof}

Now let $\gs = \gs_0 + \gO \gs_1 + o\left(\gO\right) \in \So{L}$ satisfying $n^{\gr} D_{\gr}I\left(\gs\right)_{\mu}\big|_{\scrI}=0$. Combining the two preceding lemmas we have
\begin{equation*}\label{Zeroth order structure at null infinity: intrisic operator, spacetime Thomas operator}
I\left(\gs\right)_I:= \Mtx{\gs_0\\ \gsd_0 \\ \covD_A\gs_0 \\ \gs_1 \\ -\tfrac{1}{d-2}\left(\covD^C \covD_C  -\tfrac{d-2}{2}V_2\right) \gs_0 - \tfrac{1}{2(d-2)}h^{CD}\hd_{CD}\;\gs_1}.
\end{equation*}

Since we defined null-tractors as the orthogonal space to $I\left(\gO\right)^I$, dual null-tractors are obtained by quotienting by $I\left(\gO\right)_I$ ,  $\left(\cT_{\scrI}\right)^* = (\cT)^*\big|_{\scrI} / I\left(\gO\right)$. Together with the above results this remark proves the following proposition:
\begin{Proposition}\label{Zeroth order structure at null infinity: intrisic operator, Thomas operator}
	Let $\gs_0 \in \So{L_{\scrI}}$ satisfying $\covD_a\left(\gsd_0 - \tfrac{1}{2(d-2)}h^{CD}\hd_{CD}\;\gs_0\right)=0$, it defines a section of the dual null-tractor bundle $\left(\cT_{\scrI}\right)^* = \left( \cT\right)^* \big|_{\scrI} / I\left(\gO\right)$ through
\begin{equation*}
\It\left(\gs_0\right)_I:= \Mtx{\gs_{0}\\ \gsd_0 -\tfrac{h^{CD}\hd_{CD}}{2(d-2)} \gs_0 \\ \covD_A\gs_0 \\ -\tfrac{1}{d-2}\left(\covD^C \covD_C  -\tfrac{d-2}{2}V_2\right) \gs_0}\in (\cT_{\scrI})^*
\end{equation*}
( $-\tfrac{d-2}{2}V_2 := P^h$ if $d\geq4$ and $V_2 := M$ if $d=3$).
\end{Proposition}
The appearance of an extra term on the second line is due to our conventions for dual null-tractors, see section \ref{sssection: Transformation rules}. We will call Thomas operator the operator defined by this proposition. Note that it matches the definition from \cite{herfray_asymptotic_2020}.

\paragraph{Invariant definition of null-tractors}

Let us first recall (from e.g. \cite{gover_almost_2010}) the invariant definition of the tractor bundle. Let $S^2\big|_0 (TM)^* \otimes L$ be the bundle of weighted trace-free symmetric tensor on $M$ we have a canonical injection $ S^2\big|_0 (TM)^* \otimes L  \inj J^2 L$ where $J^2 L$ is the second order jet bundle of $L\to M$. The dual tractor bundle $(\cT)^*$ of $M$ can be invariantly defined as the quotient
\begin{equation*}
(\cT)^* := \Quotient{J^2 L}{S^2\big|_0 (TM)^* \otimes L}.
\end{equation*}
By prolongation, $\bgO \in \So{L}$ defines a preferred section $J^2\gO$ of $\cT$ and one can consider the quotient $(\cT)^* / J^2\gO$. Up to now this is the restriction of this quotient at $\scrI$ that we called the (dual) null-tractor bundle. To avoid confusion, in the following discussion, we will refer to this bundle as the ``extrinsic'' (dual) null-tractor bundle.
  
We now recall from \cite{herfray_asymptotic_2020} the invariant definition of the dual null-tractor bundle $(\cT_{\scrI})^*$ as a sub-bundle of $J^2L_{\scrI}$. We will call the resulting bundle the ``intrinsic'' (dual) null-tractor bundle.

 Let $F \subset J^2L_{\scrI}$ be the sub-bundle of the second order jet bundle $J^2L_{\scrI}$ corresponding to formal solutions of 
\begin{align*}
\covD_a\left(\gsd_0 - \tfrac{1}{2(d-2)}h^{CD}\hd_{CD}\; \gs_0\right)=0.
\end{align*}
Note that lemma \ref{Zeroth order structure at null infinity: intrisic operator, nDI} ensures that these equations are conformally invariant i.e. correspond to the zeros of a well-defined operator on sections of $L_{\scrI}$. 

Let $S^2\big|_0 (T\scrI/n)^* \otimes L_{\scrI}$ be the sub-bundle of weighted trace-free symmetric tensors (whose section are, in our notation, of the form $T_{AB}$ with $h^{AB}T_{AB}=0$). We have a canonical injection $S^2\big|_0 (T\scrI/n)^* \inj F \subset J^2L_{\scrI}$. The ``intrinsic'' dual null-tractor bundle can then be invariantly defined as the quotient
\begin{equation*}
(\cT_{\scrI})^* := \Quotient{F}{S^2\big|_0 (T\scrI/n)^*}.
\end{equation*}

We will now show that the ``intrinsic'' and ``extrinsic'' null-tractor bundles are  canonically isomorphic. We in fact already wrote an explicit version of this isomorphism in the form of proposition  \ref{Zeroth order structure at null infinity: intrisic operator, Thomas operator}. This is because Thomas operator as defined in this proposition effectively yields a map of the form
\begin{equation*}
\It: \begin{array}{ccc}
\Quotient{F}{S^2\big|_0 (T\scrI/n)^*} &\to& \left. \Quotient{(\cT)^*}{I\left(\gO\right)}\right|_\scrI
\end{array}
\end{equation*}
i.e. from the intrinsic to the extrinsic (dual) null-tractor bundle. It is clear from this reasoning that for $d\geq4$ nothing in this identification depends on the choice of $\left(\MConf, \gConf_{\mu\nu}, \bgO\right)$ extending $\left(\scrI, \bh_{ab},\bn^a \right)$ (recall, however, that we assume $DI\left(\gO\right) = O\left(\gO^2\right)$). When $d=3$ the freedom in this isomorphism amounts to a choice of Mass aspect.

\section{First order structure at null infinity, the induced tractor connection}\label{First order structure at null infinity, the induced tractor connection}

\subsection{The induced tractor connection}

Let $\left(\scrI, \bh_{ab} , \bn^a\right)$ be a conformal Carroll manifold of null infinity type and let $\left(\MConf  , \gConf_{\mu\nu} , \bgO \right)$ be an asymptotically flat manifolds extending it and satisfying
\begin{equation}\label{First order structure at null infinity: Einstein's equation}
	D_{\gr}I\left(\gO\right)^I = O\left(\gO^2\right).
\end{equation}
 In the previous section we defined the null-tractor bundle $\cT_{\scrI} \to \scrI$ as the sub-bundle $I^{\perp} \subset \cT\big|_{\scrI}$ of the restriction of the spacetime tractor bundle at $\scrI$. If follows from \eqref{First order structure at null infinity: Einstein's equation} that the spacetime tractor connection induces a connection $\Dt$ on $\cT_{\scrI}$. We will now discuss the property of this induced connection.
 
A first remark is that $D_{\gr}I\left(\gO\right)^I = O\left(\gO\right)$ would have been enough to induce a connection on the tractor bundle (this is in fact a necessary condition because, as opposed to more usual hyperplane, there isn't any canonical projection on a null hyperplane). Therefore equations \eqref{First order structure at null infinity: Einstein's equation} are strictly more than is necessary to induce a tractor connection. We will see that this extra fall-off is in fact equivalent to requiring that the induced tractor connection satisfies the normality conditions from \cite{herfray_asymptotic_2020}. In fact, we will prove that all such connections can be obtained in this way.

We showed in the previous section that the null-tractor bundle is in fact intrinsic to $\left(\scrI, \bh_{ab}, \bn^a\right)$ and did not depend on the choice of extension $\left(\MConf , \gConf_{\mu\nu} , \bgO \right)$. A second remark about the induced tractor connection is that it \emph{does} depend on the choice of extension. In fact we will see that it precisely encodes the first order germ of these extensions for $d\geq4$ (respectively the second order germ for $d=3$). 

For $d=4$, we will see that the induced tractor connection can be parametrised by the asymptotic shear of null geodesic congruence, therefore the freedom in choosing a tractor connection compatible with $\left(\scrI, \bh_{ab}, \bn^a\right)$ geometrically realises the physical gravitational radiations that might be reaching null infinity.

\subsection{Normality conditions}

Let $\left(\scrI, \bh_{ab}, \bn^a\right)$ be a conformal Carroll geometry of null infinity type and let $\left(M, \gConf_{\mu\nu},\bgO\right)$ be an asymptotically flat spacetime extending it. As we just discussed, the interior normal tractor connection $D$ induces on the null-tractor bundle $\cT_{\scrI}$ a connection $\Dt$.

In this subsection we show that $\Dt$ always is a \emph{null-normal} tractor connection in the sense of \cite{herfray_asymptotic_2020}. It will follow from results of the following subsection (together with those of \cite{herfray_asymptotic_2020}) that all such connections can be obtained in this way.

\paragraph{Compatibility with the conformal Carroll geometry}

Since the spacetime normal tractor connection satisfies $D_{\gr} g_{IJ}=0$ and $D_{\gr} I\left(\gO\right)^I = O\left(\gO^2\right)$ we have
\begin{align*}
\Dt_{c} h_{IJ} &=0, & \Dt_{c} \It^I &=0.
\end{align*} 

Let $u \in \Co{\scrI}$ be well-adapted trivialisation, we recall from the previous section that it defines preferred representatives $\left( h_{ab} , n^a\right)$ in $\left[ h_{ab}, n^a \right]$ and an isomorphism $\cT_{\scrI} \xeq{u} \bbR \oplus T\scrI/n \oplus \bbR \oplus \bbR$. It follows from the definition of the normal tractor connection \eqref{Asymptotically flat spacetimes: Tractor connection, Definition} and the isomorphism \eqref{Zeroth order structure at null infinity: Null-Tractors, isomorphism} for null-tractors that
\begin{equation*}
\Dt_{c} \Phit^I \xeq{u} 
\Mtx{* & * & 0 &0 \\
	* & * & 0 &\gth^A_c \\
	* & * & \covD_c  & (du)_c \\
	* & * & 0 &\covD_c } \Mtx{\Phit^+ \\ \Phit^A \\ \Phit^u \\ \Phit^-}.
\end{equation*}

Finally since $\Dt$ is induced by the normal tractor connection $D$ we have, \begin{equation*}
\Ft^{I}{}_{Jab} = F^{I}{}_{Jab}\big|_{\scrI} 
\end{equation*} 
where $\Ft$ and $F$ are the respective curvature 2-forms. Since $X^J F^I{}_{Jab} = 0$, $\Dt$ must be torsion-free i.e. satisfies $X^J \Ft^I{}_{Jab} = 0$.

\paragraph{Compatibility with Thomas operator}
Let $\Psit_I \in \So{(\cT_{\scrI})^*}$ be a section of the dual null-tractor bundle satisfying
\begin{align}\label{First order structure at null infinity: compatibility with Thomas operator}
	\Dt_{c}\Psit_{-}&=0, & \Dt_{c}\Psit_{u} &=0,  & h^{AB }\Dt_{A}\Psit_{B}&=0,
\end{align}
then, as we shall see, $\Psit_I$ must be in the image of Thomas operator as defined by proposition \ref{Zeroth order structure at null infinity: intrisic operator, Thomas operator} i.e. $\Psit_I = \It\left(\gs_0\right)_I$ where $\bgs_0 := \bXt^I\Psit_I$.

We now prove this fact. By definition dual null-tractors are elements of  the quotient $(\cT)^*/I$ of $(\cT)^*\big|_{\scrI}$ by $I\left(\gO\right)_I$. Let $\Psi_I$ be a section of $(\cT)^*$ such that its image in $(\cT)^*/I$ coincides with $\Psit_I$ when restricted to $\scrI$. Equations \eqref{First order structure at null infinity: compatibility with Thomas operator} are then equivalent to 
\begin{align*}
	D_{c}\Psi_{-}\big|_{\scrI}&=0, &  D_{c}\Psi_{u} \big|_{\scrI}&=0,  & h^{AB}D_{A}\Psi_{B}\big|_{\scrI}&=0.
\end{align*}
Let $\gs:=\Psi_-$ with $\gs = \gs_0 + \gO \gs_1 + o\left(\gO\right)$. We can make use of the ambiguities $\Psi_I \mapsto \Psi_I + f I\left(\gO\right)_I$ and $\gs_1 \mapsto \gs_1 + g$ in the definition of $\Psi_I$ to respectively achieve 
\begin{align*}
	D_{\gr}\Psi_- \big|_{\scrI}&=0,  & g^{\mu\nu}D_{\mu}\Psi_{\nu}\big|_{\scrI} &=0.
\end{align*}
Which means that $\Psi_I\big|_{\scrI}$ is in the image of the spacetime Thomas operator,
\begin{equation*}
	\Psi_I\big|_{\scrI} = I\left(\gs\right)_I\big|_{\scrI}.
\end{equation*}
Since by hypothesis $n^c D_c \Psi_a = D_a \Psi_u =0$ we have, from results of section \ref{sssection: Thomas operator and invariant definition of null-tractors}, that $\Psit_I$ is in the image of the boundary Thomas operator: $\Psit_I = \It\left(\gs_0\right)_I$.

\paragraph{Normality conditions}

Let $D$ be the normal tractor connection of an asymptotically flat spacetime extending $\left(\scrI, \bh_{ab} ,\bn^a  \right)$, then by our assumptions:
\begin{align*}
F^I{}_{J\mu\nu} I^J &= D_{[\mu}D_{\nu]} I\left(\gO\right)^I = O\left(\gO\right).
\end{align*}
in particular $0=F^{\gr}{}_{J\mu\nu}I^J \big|_{\scrI} = W^{\gr}{}_{\gs\mu\nu}n^{\gs} \big|_{\scrI} $. Making use of the symmetry of the Weyl tensor we have both
\begin{align*}
\Ft^a{}_{bcd}n^b &= 0, &\Ft^a{}_{bcd}n^d &= 0.
\end{align*}
The second of these equations is the first normality condition from \cite{herfray_asymptotic_2020}. Since the Weyl tensor satisfies $W^{\mu}{}_{\gr\nu\gs} g^{\gr\gs}=0$ we have 
\begin{equation*}
0 = F^{\mu}{}_{\gr\nu\gs} g^{\gr\gs}\big|_{\scrI}  = F^{\mu}{}_{\gr\nu\gs} \left( h^{AB} m_A{}^{\gr} m_B{}^{\gs} + l^{\mu}n^{\nu} + n^{\mu}l^{\nu}\right)\big|_{\scrI}  = F^{\mu}{}_{A\nu B} h^{AB} \big|_{\scrI} 
\end{equation*}
implying
\begin{equation*}
\Ft^a{}_{CbD}h^{CD} = 0
\end{equation*}
which is the second normality condition from \cite{herfray_asymptotic_2020}.

\subsection{Explicit expression of the induced tractor connection}

\subsubsection{BMS expansion of asymptotically flat metric}

Let $\left( \scrI, \bh_{ab}, \bn^a\right)$ be a conformal Carroll manifold of null infinity type and let $\left( \MConf , \gConf_{\mu\nu} , \bgO \right)$ be an asymptotically flat manifolds extending it. Let us pick a set of BMS coordinates $\left(\gO , u , \gth\right)$ on a neighbourhood $U\subset \MConf$ of $\scrI$ and consider the asymptotic expansion of $g_{\mu\nu}$ in these coordinates.
\begin{Proposition}\label{First order structure at null infinity: Proposition BMS expansion}
Let $\left(\bgO, \gConf_{\mu\nu}, M\right)$ be an asymptotically flat spacetime to order $k=1$ and let $\left(\gO , u , \gth\right)$ be a set of BMS coordinates, then
\begin{equation*} 
	\begin{array}{llccc}
		g_{\mu\nu} = &\quad\; \gO^0 &\Big( & 2(du)_{\mu}(d\gO)_{\nu} + h_{AB} \left(d\gth^A\right)_{\mu}\left(d\gth^B\right)_{\nu} &\Big) \\
		
	&+ \; \;\gO^1 &\Big( &  -\; \tfrac{1}{d-2}h^{CD}\hd_{CD}\; (du)_{\mu}(du)_{\nu} + C_{AB} \left(d\gth^A\right)_{\mu}\left(d\gth^B\right)_{\nu} &\Big) \\
	
	&+ \; \; \gO^2 &\Big( & V_2 \; (du)_{\mu}(du)_{\nu} + \gb_2 \; 2(du)_{\mu}(d\gO)_{\nu} - U_2{}_A \; 2 (du)_{\mu} (d\gth^A)_{\nu} + D_{AB} \left(d\gth^A\right)_{\mu}\left(d\gth^B\right)_{\nu} &\Big)\\
	
	&+ \;\; O\left(\gO^3\right) & & &
	\end{array}
\end{equation*}	
where $C_{AB}$, $D_{AB}$ are symmetric tensors on $\scrI$ such that $h^{CD}C_{CD}=0$, $h^{CD}D_{CD} = \tfrac{1}{2}C^{CD}C_{CD}$ and satisfying the extra condition
\begin{equation}\label{First order structure at null infinity: CAB equation for d>5}
	\Cd_{AB} - \tfrac{h^{CD}\hd_{CD}}{2(d-2)} \; C_{AB} = -\tfrac{2}{d-4}R^{(h)}_{AB}\big|_0, \qquad \text{for} \; d\geq 5.
\end{equation}

The remaining coefficients are given by
\begin{align*}
	V_2 &:= -\tfrac{R^{(h)}}{(d-2)(d-3)} ,& U_2{}^A &:= -\tfrac{1}{2(d-3)} \covD_{C} C^C{}_{A},& \gb_2 &= -\tfrac{1}{16(d-2)} C^{CD}C_{CD}, &  \qquad \text{for} \; d\geq4, \\
	V_2 &:= M,& U_2{}^A &:= -N^A ,& \gb_2 &= 0, &  \qquad \text{for} \; d=3,
\end{align*} 
where the ``mass aspect'' $M$ and the ``angular momentum aspect'' $N^A$ are tensors on $\scrI$.
\end{Proposition}
\begin{proof}
	A proof of this classical result (see \cite{madler_bondi-sachs_2016} for a review) is given in appendix \ref{section: Appendix, BMS expansion and Einstein equations}.
\end{proof}

Therefore for $d\geq4$ the asymptotic freedom in $\left(\MConf,\hConf_{\mu\nu}, \bgO\right)$ is given, in BMS coordinates by a choice of symmetric trace-free tensor $C_{AB}$ satisfying \eqref{First order structure at null infinity: CAB equation for d>5}. For $d=4$ this tensor does not have to satisfy any differential equation and $\Cd_{AB}$ is the Bondi News tensor. Finally for $d=3$ this tensor is identically zero and the asymptotic freedom in the BMS expansion is a choice of ``mass'' and ``angular momentum aspects'' , $M$ and $N^A$, which are tensors on $\scrI$. We will see that these tensors can be seen to explicitly parametrize the induced tractor connection on the null-tractors bundle.

For $d\geq4$ there is another trace-free tensor in this expansion $D_{AB}\big|_0$. For $d=4$ and if one assumes both Einstein's equations to one order higher $D_{\gr}I^I = O\left(\gO^3\right)$ and enough differentiability so that the peeling theorem holds this tensor must vanish. However we will not need to assume this here and the tractor connection will in fact always ignore $D_{AB}$.

By proposition \ref{Proposition: Asymptotically flat spacetimes 2}, an asymptotically flat spacetime to order $k=1$ satisfies $D_{\gr} I\left(\gO\right)^{\mu} = O\left(\gO^2\right)$, $D_{\gr} I\left(\gO\right)^{-} = O\left(\gO\right)$. Requiring instead, as in proposition \ref{Proposition: Asymptotically flat spacetimes ALT}, $D_{\gr}I\left(\gO\right)^I = O\left(\gO^2\right)$ only makes a difference in dimension $d=3$:
\begin{Proposition}\label{First order structure at null infinity: Proposition BMS expansion 2}
	Let $\left(\MConf, \gConf_{\mu\nu}, \bgO\right)$ be an asymptotically flat spacetime to order $k=1$ such that $D_{\gr}I\left(\gO\right)^I = O\left(\gO^2\right)$. Then proposition \ref{First order structure at null infinity: Proposition BMS expansion} is unchanged for $d\geq4$. For $d=3$ we have the extra equations,
	\begin{align*}
	\Md + \gTh M - \covD^C \covD_C \gTh&=0, & \Nd_A + \tfrac{\gTh}{2}N_A &= \tfrac{1}{2}\covD_A M,
	\end{align*}
	where $\gTh := \tfrac{1}{d-2}h^{CD}\hd_{CD}$. These are the so-called ``conservation equations'' for the mass and angular momentum aspects.
\end{Proposition}
\begin{proof}
	See appendix \ref{section: Appendix, BMS expansion and Einstein equations}.
\end{proof}

\subsubsection{The induced tractor connection on null-tractor in BMS coordinates}

Making use of results from appendix \ref{section: Appendix, BMS expansion of geometrical quantities} one directly derives the expression of the normal tractor connection \eqref{Asymptotically flat spacetimes: Tractor connection, Definition} in terms of the BMS expansion.

\begin{Proposition}
Let $\left(\MConf ,\gConf_{\mu\nu}, \bgO\right)$ be an asymptotically flat spacetime to order $k=1$. Let $\left(\gO, u, \gth\right)$ be a set of BMS coordinates. Then the restriction of the normal tractor connection	at $\scrI$ is 
\begin{equation*}
D_c \Phi^I\big|_{\scrI} \xeq{u}
\Mtx{ \covD_c & -(du)_c & -\gth_{cB} & 0 &0 \\[0.4em]
\tfrac{1}{2} \covD_c \gTh  & -\tfrac{\gTh}{2} (du)_c + \covD_c& -\tfrac{\gTh}{2} \gth_{cB} &0 & 0 \\[0.4em]
-\xi^A_c & \tfrac{1}{2}C^A{}_C \; \gth^C_c  & \gd^A{}_B\left(\tfrac{\gTh}{2}(du)_c + \covD_c\right)  & \tfrac{\gTh}{2}\gth^A_c & \gth^A_c \\[0.4em]
-\psi_c & 0 & -\tfrac{1}{2}C{}_{BC} \; \gth^C_c& \tfrac{\gTh}{2} (du)_c +\covD_c & (du)_c\\[0.4em]
0 & \psi_c & \xi_{cB} & -\tfrac{1}{2}\covD_c \gTh & \covD_c}
 \Mtx{\Phi^+ \\[0.4em] \Phi^{\gO} \\[0.4em] \Phi^{B} \\[0.4em] \Phi^u \\[0.4em] \Phi^-} 
\end{equation*}
where $\gTh:=\tfrac{1}{d-2}h^{AB}\hd_{AB}$ and 
\begin{align*}
\xi_{Ac} &:= \tfrac{1}{2} \left(\Cd_{AC} + h_{AC} V_2\right) \;\gth^C_c +\tfrac{1}{2} \covD_A \gTh \; (du)_c, & \psi_c & := U_{2C} \; \gth^C_c - \tfrac{1}{2} V_2 \;(du)_c.
\end{align*}
\end{Proposition}
Making us of the isomorphism \eqref{Zeroth order structure at null infinity: Null-Tractors, isomorphism} for null-tractors we obtain an explicit expression for the induced tractor connection.
\begin{Proposition}\label{First order structure at null infinity: Proposition Induced Tractor connection}
	Let $\left(\MConf,\gConf_{\mu\nu}, \bgO \right)$ be an asymptotically flat spacetime to order $k=1$. Let $\left(\gO, u, \gth\right)$ be a set of BMS coordinates. Then the tractor connection induced on $\cT_{\scrI}$ is  
	\begin{equation*}
	\Dt_c \Phit^I \xeq{u} 
	\Mtx{ -\tfrac{\gTh}{2}(du)_c + \covD_c & -\gth_{cB} & 0 & 0 \\[0.4em]
		-\xi^A{}_{c} +\tfrac{\gTh}{4}C^A{}_{C}\;\gth^C_c & \gd^A{}_B\left( \tfrac{\gTh}{2} (du)_c + \covD_c\right)& 0 & \gth^A_c \\[0.4em]
		-\psi_c & -\tfrac{1}{2}C_{BC} \;\gth^C_c & \covD_c  & (du)_c \\[0.4em]
		0 & \xi_{Bc} -\tfrac{\gTh}{4}C_{BC}\;\gth^C_c & 0 & \tfrac{\gTh}{2}(du)_c + \covD_c }
	\Mtx{\Phit^+ \\[0.4em] \Phit^{B} \\[0.4em] \Phit^u \\[0.4em] \Phit^-}.
	\end{equation*}
where
\begin{align*}
\gTh&:=\tfrac{1}{d-2}h^{AB}\hd_{AB}, &\xi_{Ac} &:= \tfrac{1}{2} \left(\Cd_{AC} + h_{AC} V_2\right) \;\gth^C_c +\tfrac{1}{2} \covD_A \gTh \; (du)_c, & \psi_c & := U_{2C} \; \gth^C_c - \tfrac{1}{2} V_2 \;(du)_c,
\end{align*}
and $V_2$, $U_2^A$ are given by proposition \ref{First order structure at null infinity: Proposition BMS expansion}.
\end{Proposition}
For future use this is also useful to have the expression of the induced tractor connection acting on dual null-tractors:
\begin{equation*}
\Dt_c \Psit_I \xeq{u} 
\Mtx{ -\tfrac{\gTh}{2}(du)_c + \covD_c & -(du)_c&-\gth_{cB} & 0  \\[0.4em]
	0 & \covD_c& 0 & 0 \\[0.4em]
	-\xi_{Bc}+ \tfrac{\gTh}{4} C_{BC}\; \gth^C_c & \tfrac{1}{2}C{}_{BC} \;\gth^C_c & -\tfrac{\gTh}{2}(du)_c + \covD_c  & \gth_{Bc} \\[0.4em]
	0 & \psi_c & \xi^B{}_c -\tfrac{\gTh}{4}C^B{}_C\;\gth^C_c & \tfrac{\gTh}{2}(du)_c + \covD_c }
\Mtx{\Psit_- \\[0.4em] \Psit_{u} \\[0.4em] \Psit_B \\[0.4em] \Psit_+}.
\end{equation*}

\subsubsection{Discussion}

Note that the expressions in proposition \ref{First order structure at null infinity: Proposition Induced Tractor connection} match those of \cite{herfray_asymptotic_2020} if one works in the gauge where $\gTh := \tfrac{1}{d-2}h^{CD}\hd_{CD}=0$. It was shown \cite{herfray_asymptotic_2020}, given a conformal Carroll structure (of null infinity type) all compatible null-normal tractor connections are of this form. Therefore, given a conformal Carroll manifold $\left(\scrI, \bh_{ab} , \bn^a\right)$, null-normal tractor connections can be obtained by choosing a suitable extension $\left(\bgO, \gConf_{\mu\nu}\right)$ and restricting the related normal tractor connection. 

Together with proposition \ref{First order structure at null infinity: Proposition BMS expansion}, these remarks imply that choices $\left( \scrI, \bh_{ab}, \bn^a , \Dt\right)$ of conformal Carroll geometry together with a compatible null-normal tractor connection are in one-to-one correspondence with first order germs of asymptotically flat manifolds for $d\geq4$ and second order germs of asymptotically flat manifolds for $d\geq3$. For $d=4$, assuming the peeling would impose $D_{AB}\big|_0=0$ and $\left( \scrI, \bh_{ab} ,\bn^a , \Dt , \right)$ would in fact be equivalent to second order germs of asymptotically flat manifolds as well.

\section{The tractor curvature and physical interpretations}

From \eqref{Asymptotically flat spacetimes: Tractor curvature, Definition} and $\Ft^I{}_{Jab} = F^I{}_{Jab} \big|_{\scrI}$, the tractor curvature of the induced tractor connection  $\Dt$ on $\cT_{\scrI}$ is
\begin{equation}\label{The tractor curvature and physical interpretation : general tractor curvature}
\Ft^I{}_{Jcd} \xeq{u} 
\Mtx{0 & 0 & 0& 0 \\[0.4em] 
C^{(0)}_{cd}{}^A - \tfrac{\gTh}{2}W^{(0)}{}_{\gO}{}^{A}{}_{cd} & W^{(0)}{}^A{}_{Bcd} & 0 & 0 \\[0.4em]
C^{(0)}_{cd}{}^u & W^{(0)}{}_\gO{}_{Bcd} & 0 & 0\\[0.4em]
0 & -C^{(0)}_{cd B} +\tfrac{\gTh}{2}W^{(0)}{}_\gO{}_{Bcd}&0 & 0  }
\end{equation}
where $W^{(0)}_{\mu}{}^{\nu}{}_{\gr\gs} := W_{\mu}{}^{\nu}{}_{\gr\gs} \big|_{\scrI}$ and $C^{(0)}_{\gr\gs}{}^{\mu} := C_{\gr\gs}{}^{\mu} \big|_{\scrI}$ are the restriction of the Weyl and Cotton tensors of $g_{\mu\nu}$ to $\scrI$.

It follows from the interpretation of $\Dt$ as a Cartan connection modelled on
 \begin{equation*}
	\scrI^{d-1} := \Quotient{\ISO\left(d-1,1\right)}{\left( \bbR \times ISO\left(d-2\right)\right) \ltimes \bbR^{d-1}}
\end{equation*}
(which is a realisation of null infinity as an homogeneous space) that its curvature vanishes if and only if there exists a well-adapted trivialisation $u \in \Co{\scrI}$ such that the corresponding asymptotic shear vanishes (respectively, for $d=3$, such that the corresponding mass and angular momentum aspects vanish), see \cite{herfray_asymptotic_2020} for more details.

Depending on the dimension, the precise physical content of this curvature tensor greatly differs.

\subsection{\texorpdfstring{The tractor curvature for $d=3$ and the ``conservation equations''}{The tractor curvature for d=3 and the ``conservation equations''}}

For $d=3$ the tractor curvature of the induced tractor connection $\Dt$ is
\begin{equation*}
\Ft^I{}_{Jab} = \Mtx{0 & 0 & 0 & 0 \\[0.4em]
-\left(\Md + \gTh M - \covD^C\covD_C \gTh \right) h_{BD} & 0 & 0 & 0 \\[0.4em]
2\Nd_D +\gTh N_D -\covD_D M & 0 & 0& 0 \\[0.4em]
0 & \left(\Md +\gTh M -\covD^C\covD_C \gTh \right) h_{BD} & 0 & 0} (du)_{[c} \gth^D_{d]}
\end{equation*}
where $\gTh := \tfrac{1}{d-2}h^{CD}\hd_{CD}$. One sees that the curvature of the tractor connection encodes the ``conservation equations'' for the mass and angular momentum aspect. As was discussed in proposition \ref{First order structure at null infinity: Proposition BMS expansion 2} if one requires that the asymptotically flat spacetime satisfies $D_{\gr} I\left(\gO\right) =O\left(\gO^2\right)$ then it vanishes identically.

In other terms, when the ``conservation equations'' hold the mass and angular momentum aspects are coordinates expression parametrizing a flat tractor connection. From the previous discussion and results from \cite{herfray_asymptotic_2020} all flat null-normal tractor connection can be obtained in this way.

\subsection{\texorpdfstring{The tractor curvature for $d=4$ and gravitational radiations}{The tractor curvature for d=4 and gravitational radiations}}

For $d=4$ the tractor curvature of the induced tractor connection $\Dt$ is
\begin{equation*}
\Ft^I{}_{Jab} = \Mtx{0 & 0 & 0 & 0 \\[0.4em]
-\eps^{AE}K_E \;\eps_{CD}\gth_c{}^C \gth_d{}^D -K^A{}_D \;(du)_{[c} \gth_{d]}{}^D & 0 & 0 & 0 \\[0.4em]
\tfrac{1}{2}K \;\eps_{CD}\gth_c{}^C \gth_d{}^D + 2 K_D \;(du)_{[c} \gth_{d]}{}^D   &0 & 0& 0 \\[0.4em]
	0 & \eps_B{}^{E}K_E \;\eps_{CD}\gth_c{}^C \gth_d{}^D +K_{BD} \;(du)_{[c} \gth_{d]}{}^D & 0 & 0}
\end{equation*}
where
\begin{align*}
K_{AB} &:= \Cdd_{AB} +\left(\tfrac{\gTh}{2}C_{AB}\right)^. - \covD_A\covD_B\big|_0 \gTh, &
K_A & := \tfrac{1}{2} \covD^C \left( \Cd_{AC} +\tfrac{\gTh}{2}C_{AC}\right) + \tfrac{1}{4} \covD_A R^{(h)},
\end{align*}
\begin{align*}
K & := \eps^{CD} \left( \covD_C \covD^E C_{DE} + \tfrac{1}{2} C^E{}_{C}\Cd_{DE}   \right).
\end{align*}
and $\gTh:=\tfrac{1}{d-2}h^{CD}\hd_{CD}$. Comparing with \eqref{The tractor curvature and physical interpretation : general tractor curvature} one sees that the restriction of the Weyl tensor $W^{\mu}{}_{\nu cd}\big|_{\scrI}$ vanishes identically. Note that this does not implies the peeling since $W_{\gO\mu\gO\nu}\big|_{\scrI}$ might still be non-vanishing here. The tractor curvature of the induced tractor connection is therefore entirely parametrized by the restriction of the Cotton tensor $C_{cd}{}^{\mu}\big|_{\scrI}$.

Let us introduce the gravitational tensor $K{}_{\mu\nu cd} := \gO^{-1} W_{\mu\nu cd}\big|_{\scrI}$. Making use of the identity \eqref{Asymptotically flat spacetimes: Cotton tensor, identities} for $d=4$, $C_{\gr\gs}{}_{\mu} = \covD_{\nu} W^{\nu}{}_{\mu\gr\gs}$ and  we have
\begin{equation*}
C_{cd}{}^{\mu}\big|_{\scrI} =  (d\gO)_{\gr}\;  K^{\gr}{}^{\mu}{}_{ cd} = -K^{\mu}{}_{\gr cd} \; n^{\gr}
\end{equation*}
and therefore the data of the boundary Cotton tensor is equivalent to the Newman-Penrose coefficients
\begin{align*}
\Psi_4^0 & := K{}_{\mu \gr cd} \; \mb^{\mu}\; n^{\gr} \; n^{c} \; \mb^d, & \Psi_3^0 & := K_{\mu \gr cd} \; l^{\mu} \; n^{\gr} \; n^{c} \; \mb^d ,  & Im\left(\Psi_2^0\right) &:= \tfrac{1}{2} K_{\mu \gr cd} \; l^{\mu}\; n^{\gr} \; m^c \; \mb^d.
\end{align*}
These are well-known to encode the presence of gravitational radiations.

In others terms, while the asymptotic shear $C_{AB}$ is a coordinate expression parametrizing the induced tractor connection, the Newman-Penrose coefficients
 $\Psi_4^0$, $\Psi_3^0$, $Im\left(\Psi^0_2\right)$ are coordinates for its curvature. 
This fleshes our claim that, in four dimensions, gravitational radiations are encoded in the curvature of the induced tractor connection.

 The tractor connection itself encodes the extra information given by the zero mode of the asymptotic shear and news. These boundary degrees of freedom are dynamical with equations of motion given by the invariant equation
\begin{equation*}
\Ft^I{}_{Jcd} = J^I{}_{Jcd}
\end{equation*}
where $J^I_{Jcd}$ is a source term describing the flux of gravitational radiations. This suggests that the dynamics of radiative degrees of freedom at null infinity might be described by an effective Chern-Simon action for the tractor connection coupled with external sources, see \cite{nguyen_effective_2021} for a discussion on such variational principle (however not including sources).

As we already discussed at the beginning of this section, in the absence of gravitational radiation i.e. whenever the above curvature vanishes, the induced tractor connection is flat  and the corresponding ``asymptotic shear'' $C_{AB}$ is ``pure gauge'': there exists a choice of well-adapted trivialisation $u$ such that it vanishes.

\subsection{\texorpdfstring{The tractor curvature for $d\geq5$ and zero modes}{The tractor curvature for d>=5 and zero modes}}

For $d\geq5$ the tractor curvature of the induced tractor connection $\Dt$ is
\begin{equation*}
\Ft^I{}_{Jab} = \Mtx{0 & 0 & 0 & 0 \\[0.4em]
	-C^{(h)}{}_{CD}{}^A & W^{(h)}{}^A{}_{BCD} & 0 & 0 \\[0.4em]
	\tfrac{1}{d-3}\covD_{[C}\covD^E C_{D]E} - C_{B[C} P^{(h)}{}^B{}_{D]} & -\covD_{C}C_{DB} -\tfrac{1}{d-3}\covD^{E}C_{EC} h_{DB} & 0& 0 \\[0.4em]
	0 & -C^{(h)}{}_{CDB} & 0 & 0} \gth_{[c}{}^C \gth_{d]}{}^D
\end{equation*}
where $W^{(h)}{}^A{}_{BCD}$, $C^{(h)}{}_{AB}{}^{C}$ and $P^{(h)}{}_{AB}$ are respectively the Weyl, Cotton and Schouten tensors of the $(d-2)$-dimensional conformal metric $\bh_{AB}$.

From proposition \ref{First order structure at null infinity: Proposition BMS expansion} evolution of the asymptotic shear $C_{AB}$ along the null direction is completely determined by the conformal geometry $\bh_{AB}$. The value of the tractor connection on any section of $\scrI \to \gS$ therefore completely determines it. Therefore in these dimensions there is no freedom in the dynamics of the induced tractor connection. This is in line with the well known fact that in dimension $d\geq5$ gravitational radiations are encoded in sub-leading terms of the BMS expansion and not in the asymptotic shear.

Under the stronger assumptions of \cite{hollands_bms_2017}, in particular assuming Einstein's vacuum equations to infinite order $DI^I =0$,  this curvature must vanish. In this sense the situation for $d\geq5$ is similar to $d=3$. 

\appendix

\section{Appendix: BMS expansion and Einstein equations}\label{section: Appendix, BMS expansion and Einstein equations}

We here gather facts about the expression of Einstein's equations to lowest order in BMS coordinates. This is has been well studied in dimension three and four, see \cite{bondi_gravitational_1962,sachs_gravitational_1962,tafel_comparison_2000,barnich_aspects_2010,barnich_finite_2016,madler_bondi-sachs_2016}, but higher dimensions are not as well covered, see however \cite{tanabe_asymptotic_2011,kapec_higher-dimensional_2017,hollands_bms_2017,campoleoni_asymptotic_2018-1,satishchandran_asymptotic_2019,campoleoni_asymptotic_2020}. Our conventions mainly follows those of \cite{barnich_aspects_2010}.

\subsection{BMS expansion}

Let $\left(\MConf, \gConf_{\mu\nu}, \bgO\right)$ be an asymptotically simple manifolds. Let $\left(u, \gO, \gth \right)$ be a choice of BMS coordinates, we have:
\begin{equation}\label{Appendix: Well-adapted trivialisation, BMS coordinates}
	g_{\mu\nu} = \gO^3 e^{2\gb}  V (du)_{\mu}(du)_{\nu} + e^{2\gb} 2 (du)_{\mu} (d\gO)_{\nu} + H_{AB}\left((d\gth)_{\mu}{}^{A} -U^A (du)_{\mu} \right)\left( (d\gth)_{\nu}{}^{B} -U^B (du)_{\nu} \right)
\end{equation}
by definition of asymptotically simple manifolds and BMS coordinates one has $\gb \big|_{\scrI}=0$, $\gO^3 V\big|_{\scrI}=0$, $U^A\big|_{\scrI}=0$ and $H_{AB}\big|_{\scrI} = h_{AB}$ and $\parD_{\gO}det(H_{AB}) =  0$. Assuming that both $g_{\mu\nu}$ and $\gO$ are of class $\Cc^3$, one has the asymptotic expansions:
\begin{align*}
	\gb &= \gO \gb_1 + \gO^2 \gb_2 + O\left(\gO^3\right), & \gO^3 V &= \gO V_1 + \gO^2 V_2 +  O\left(\gO^3\right), \\
	U^A &= \gO U_1^A + \gO^2 U_2^A + O\left(\gO^3\right), & H_{AB} &= h_{AB} + \gO C_{AB} +\gO^2 D_{AB} +  O\left(\gO^3\right).
\end{align*}
where $h^{AB}C_{AB} =0$, $h^{AB}D_{AB} = \tfrac{1}{2} C^{CD} C_{CD}$. As a convention, all upper Latin indices from the beginning of the alphabet are raised and lowered with $h_{AB}$.

One therefore has the matrix expansion
{\scriptsize
\vspace{-0.3cm}	
	  \begin{adjustwidth}{-0,75cm}{}
\begin{align*}
\Mtx{g_{uu} & g_{u\gO} & g_{uB} \\
g_{\gO u} & g_{\gO \gO} & g_{\gO B}\\
g_{Au}& g_{A \gO} & g_{AB}}
 &=
  \Mtx{0 & 1 & 0 \\ 1& 0 &0 \\ 0 & 0 & h_{AB} } + \gO \Mtx{V_1 & 2\gb_1 & -U_1{}_B \\ 2\gb_1 & 0 & 0 \\  -U_1{}_A & 0 & C_{AB}} 
  + \gO^2 \Mtx{V_2 + 2 \gb_1 V_1 & 2\gb_2 + 2 (\gb_1)^2& -U_2{}_B -C_{BC}U_1^C \\ 2\gb_2 + 2 (\gb_1)^2 & 0 & 0 \\  -U_2{}_A -C_{AC}U_1^C & 0 & D_{AB}} + O\left(\gO^3\right)
\end{align*}
\end{adjustwidth}}

\subsection{Extrinsic curvature}

In the BMS coordinates system $\left(u, \gO, \gth\right)$, one has the following expansion for the Christoffel symbols:
{\small \begin{flalign*}
&\gC^{\mu}{}_{\nu u} =  \Mtx{ - \tfrac{1}{2}V_1 & 0 & \tfrac{1}{2}U_1{}_B \\[0.2em] 0 & \tfrac{1}{2}V_1  & 0 \\[0.2em] 0 & -\tfrac{1}{2}U_1{}^C & \tfrac{1}{2}h^{CD} \hd_{DB}} +
O\left(\gO\right).
\end{flalign*}}
The extrinsic curvature is $K^a{}_b := \covD_b n^a\big|_{\scrI} = \gC^a{}_{bu} \big|_{\scrI}$. Defining the trace-free extrinsic curvature as $\Kr^a{}_b := K^a{}_b - \gd^a{}_b \tfrac{K^c{}_c}{d-1}$ one therefore has
\begin{align*}
\Mtx{\Kr^u{}_u & \Kr^u{}_B \\ \Kr^A{}_u & \Kr^A{}_B } &= \Mtx{ \gk & \tfrac{1}{2} U_1{}_B \\ 0 & h^{CD}\left( \tfrac{1}{2}\hd_{DB}\big|_0 -\tfrac{\gk}{d-2} h_{DB} \right)},  & \tfrac{1}{d-1} K^a{}_a & = -\tfrac{1}{2}V_1-\gk 
\end{align*}
where $\gk := \tfrac{1}{2}\left(\tfrac{1}{d-2}h^{CD}\hd_{CD} +V_1\right)$ and $\big|_0$ indicates trace-free part. 

Vanishing of the trace-free extrinsic curvature (which is a conformally invariant equation) is equivalent to
\begin{align}
V_1 & = -\tfrac{1}{d-2}h^{CD}\hd_{CD},& U_1{}_A & =0, & \hd_{AB} = \tfrac{h^{CD}\hd_{CD}}{d-2}\; h_{AB}.
\end{align}

\subsection{Einstein equations to lower orders}

Proposition \ref{Proposition: Asymptotically flat spacetimes 2} asserts that Einstein equations $\Rt_{\mu\nu} - \tfrac{1}{2}\Rt \ggt_{\mu\nu} = O\left(\gO^{k-1}\right)$ ($k\geq1$) are equivalent to $D_{\gr}I\left(\gO\right)^{\mu} = O\left(\gO^{k}\right)$.

\subsubsection{\texorpdfstring{Einstein equation to lowest order: $D_{\gr}I\left(\gO\right)^{\mu} = O\left(\gO^{1}\right)$}{Einstein equation to lowest order: DI = O(Omega)}}\label{Appendix, BMS expansion: Vanishing extrinsic curvature}

As was discussed in section \ref{sssection: Einstein's equations at lowest order}, $D_c{}I\left(\gO\right)^a = O\left(\gO\right)$, is equivalent to the vanishing of the trace-free extrinsic curvature and given in BMS coordinates by $\eqref{Appendix, BMS expansion: Vanishing extrinsic curvature}$. 

A direct computation shows that requiring the Einstein tensor to be finite at the boundary, equivalently $D_{\gr}I\left(\gO\right)^{\mu} = O\left(\gO\right)$, is equivalent to 
\begin{align}\label{Appendix, BMS expansion: Einstein equation to lowest order}
V_1 & = -\tfrac{1}{d-2}h^{CD}\hd_{CD},& U_1{}_A & =0, & \hd_{AB} = \tfrac{h^{CD}\hd_{CD}}{d-2}\; h_{AB}, & &\gb_1 &=0.
\end{align}

\subsubsection{\texorpdfstring{Einstein equation to lowest order: $D_{\gr}I\left(\gO\right)^{\mu} = O\left(\gO^{2}\right)$}{Einstein equation to lowest order: DI = O(Omega square)}}

Let us assume Einstein equations to lowest order \eqref{Appendix, BMS expansion: Einstein equation to lowest order}. To next order, Einstein equations $D_{\gr}I\left(\gO\right)^{\mu} = O\left(\gO^{2}\right)$ are found to be equivalent to
\begin{align}\label{Appendix, BMS expansion: Einstein equation to second lowest order}
(d-3)V_2 &= -\tfrac{1}{d-2} R^h, & (d-3) U_{2}{}_A &= -\tfrac{1}{2} \covD_{C} C^C{}_A, \\
\tfrac{d-4}{2}\left( \Cd_{AB} -\tfrac{h^{CD}\hd_{CD}}{2(d-2)} \; C_{AB} \right) &= -R^h_{AB}\big|_0, & \gb_2& = -\tfrac{1}{16(d-2)} C^{CD}C_{CD}. \nonumber
\end{align}

Finally one can show that for $d\geq4$ equations \eqref{Appendix, BMS expansion: Einstein equation to second lowest order} are in fact equivalent to $D_{\gr}I\left(\gO\right)^{I} = O\left(\gO^{2}\right)$ (which is formally stronger since it also requires $D_{\gr}I\left(\gO\right)^{-} =  O\left(\gO^{2}\right)$). For $d=3$, however, $D_{\gr}I\left(\gO\right)^{I} = O\left(\gO^{2}\right)$ implies, on top of \eqref{Appendix, BMS expansion: Einstein equation to second lowest order},
\begin{align*}
\Vd_2 +\gTh V_2 - \covD^C \covD_C \gTh &=0 & \Ud_2{}_A +\tfrac{\gTh}{2}U_{2}{}_A &= -\tfrac{1}{2}\covD_A V_2,
\end{align*}
where $\gTh:=\tfrac{1}{d-2}h^{CD}\hd_{CD}$. These are the so-called conservation equations for the mass and angular momentum aspects:
\begin{align*}
M&:=V_2, & N_A&:=U_2{}_A.
\end{align*}

\section{Appendix: BMS expansion of geometrical quantities associated to the unphysical metric}\label{section: Appendix, BMS expansion of geometrical quantities}

We here assume $D_{\gr}I\left(\gO\right)^{\mu} = O\left(\gO^{2}\right)$, from the previous appendix or proposition \ref{First order structure at null infinity: Proposition BMS expansion} one has the matrix expansion
{\small 
\begin{align*}
\Mtx{g_{uu} & g_{u\gO} & g_{uB} \\
	g_{\gO u} & g_{\gO \gO} & g_{\gO B}\\
	g_{Au}& g_{A \gO} & g_{AB}}
&=
\Mtx{0 & 1 & 0 \\ 1& 0 &0 \\ 0 & 0 & h_{AB} } + \gO \Mtx{ -\gTh & 0 & 0 \\0 & 0 & 0 \\  0 & 0 & C_{AB}} + \gO^2 \Mtx{V_2 & 2\gb_2 & -U_2{}_B \\ 2\gb_2 & 0 & 0 \\  -U_2{}_A & 0 & D_{AB}} + O\left(\gO^3\right)
\end{align*}}\\ with $V_2$, $U_2{}_A$, $\gb_2$, $\Cd_{AB}$ satisfying \eqref{Appendix, BMS expansion: Einstein equation to second lowest order} and $h^{CD}C_{CD}=0$, $h^{CD}D_{CD} =\tfrac{1}{2}D^{CD}D_{CD}$. Everywhere in this section, $\gTh:=\tfrac{1}{d-2}h^{CD}\hd_{CD}$.
\subsection{Christoffel symbols}

In the BMS coordinates system $\left(u, \gO, \gth\right)$ and assuming \eqref{Appendix, BMS expansion: Einstein equation to second lowest order} one has the following expansion for the Christoffel symbols:
{\footnotesize 
\begin{flalign*}
\gC^{\mu}{}_{\nu u} &=  \Mtx{ \tfrac{1}{2}\gTh & 0 & 0 \\[0.2em] 0 & -\tfrac{1}{2}\gTh  & 0 \\[0.2em] 0 & 0 & \tfrac{1}{2}\gTh } 
+ \gO 
\Mtx{-V_2&0  & U_2{}_B\\[0.2em]
	-\tfrac{1}{2}\gThd +\tfrac{1}{2}\gTh^2 & V_2 & -\tfrac{1}{2}\covD_B \gTh \\[0.2em]
	\tfrac{1}{2}\covD^C \gTh & -U_2{}^C & \tfrac{1}{2}h^{CD}\Cd_{DB} -\tfrac{ \gTh }{2}C^{C}{}_B }
+ O\left(\gO^2\right), &
\end{flalign*}

\begin{flalign*}
\gC^{\mu}{}_{\nu \gO} &=  \Mtx{ 0 & 0 & 0 \\[0.2em] -\tfrac{1}{2}\gTh  & 0 & 0\\[0.2em] 0& 0& \tfrac{1}{2} C^C{}_B}
+ O\left(\gO\right),&
\end{flalign*}

\begin{adjustwidth}{-0,0cm}{0,0cm}
\begin{flalign*}
&\gC^{\mu}{}_{\nu A} =  \Mtx{ 0 & 0 & -\tfrac{1}{2}C_{AB} \\[0.2em] 0 & 0  & -\tfrac{\gTh}{2}h_{AB} \\[0.2em] \tfrac{\gTh}{2}\gd^C_{A} & \tfrac{1}{2}C^C{}_A & \gC_h{}^C{}_{BA}}
 + \gO 
\Mtx{U_2{}_A &0  & -D_{AB}  \\[0.2em]
	-\tfrac{1}{2}\covD_A\gTh & -U_2{}_A & -\tfrac{1}{2}\Cd_{AB} + \tfrac{\gTh}{2}C_{AB} \\[0.2em]
	\tfrac{1}{2}h^{CD}\left( \Cd_{DA} - \gTh C_{DA} \right)& D^C{}_A -\tfrac{1}{2}C^{CD} C_{DA} &  \covD_{(A}C_{B)}{}^C - \tfrac{1}{2}\covD^{C}C_{AB}  } + O\left(\gO^2\right).&
\end{flalign*}
\end{adjustwidth}}

\subsection{Schouten tensor}

The Ricci tensor in BMS coordinates reads
{\small \begin{equation*}
	\Mtx{R_{uu} & R_{u\gO} & R_{uB} \\ 
	R_{\gO u} & R_{\gO\gO} & R_{\gO B} \\
R_{A u} & R_{A \gO} & R_{AB}} = \Mtx{ -\tfrac{d-2}{2}\gThd & V_2 & -\tfrac{d-2}{2} \covD_B \gTh \\[0.3em]
 * & -\tfrac{1}{2} C^D{}_C C^C{}_D & -U_2{}_{B} + \tfrac{1}{2}\covD^{C} C_{CB} \\[0.3em]
 * & * & R_{AB} - \Cd_{AB} -\gTh\tfrac{(d-4)}{4}C_{AB} } + O\left(\gO\right).
\end{equation*}}\\
While the scalar curvature is
\begin{equation*}
R = R^h + 2 V_2 + O\left(\gO\right).
\end{equation*}
From which one obtains the Schouten tensor for $d\geq4$
{\small \begin{equation*}
\Mtx{P_{uu} & P_{u\gO} & P_{uB} \\ 
	P_{\gO u} & P_{\gO\gO} & P_{\gO B} \\
	P_{A u} & P_{A \gO} & P_{AB}} = \Mtx{ -\tfrac{1}{2}\gThd & -\tfrac{1}{d-2}P^h & -\tfrac{1}{2} \covD_B \gTh \\[0.3em]
	* & -\tfrac{1}{2} C^{CD} C_{CD} & \tfrac{1}{2(d-3)} \covD^C C_{CB} \\[0.3em]
	* & * & -\tfrac{1}{2}\Cd_{AB} +\tfrac{1}{d-2}h_{AB}P^h } + O\left(\gO\right),
\end{equation*}}\\and for $d=3$
{\small \begin{equation*}
\Mtx{P_{uu} & P_{u\gO} & P_{uB} \\ 
	P_{\gO u} & P_{\gO\gO} & P_{\gO B} \\
	P_{A u} & P_{A \gO} & P_{AB}} = \Mtx{ -\tfrac{1}{2}\gThd & \tfrac{1}{2}M & -\tfrac{1}{2} \covD_B \gTh \\[0.3em]
	* &0& N_B \\[0.3em]
	* & * & -\tfrac{1}{2}h_{AB} M } + O\left(\gO\right).
\end{equation*}}

\section{Appendix: Proof of proposition \ref{Zeroth order structure at null infinity: Proposition Well-adapted trivialisation, Main Proposition}}\label{section: Appendix, Proof of BMS-transformation Proposition}

Let $\left(\MConf ,\gConf_{\mu\nu},\bgO\right)$ be an asymptotically flat spacetime to order $k=1$ and let $\left( u ,\gO, \gth\right)$ be set of BMS coordinates. From proposition \ref{First order structure at null infinity: Proposition BMS expansion} one has
\begin{equation*} 
\begin{array}{llccl}
g_{\mu\nu} = &\quad\; \gO^0 &\Big( & 2(du)_{\mu}(d\gO)_{\nu} + h_{AB} \left(d\gth^A\right)_{\mu}\left(d\gth^B\right)_{\nu} &\Big) \\ 

&+ \; \;\gO^1 &\Big( &  -\; \gTh \; (du)_{\mu}(du)_{\nu} + C_{AB} \left(d\gth^A\right)_{\mu}\left(d\gth^B\right)_{\nu} &\Big) + \;\; O\left(\gO^2\right)
\end{array}
\end{equation*}	
where $\gTh := \tfrac{1}{d-2}h^{CD}\hd_{CD}$, $h^{CD}C_{CD}=0$.

Let $\left(\uh , \gOh= \go \gO, \gthh\right)$ be another set of coordinates asymptotically defined as
\begin{align*}
\uh &:= u_0 + \gO u_1 + O\left(\gO^2\right),&
\go &:=  \go_0 + \gO \go_1 + O\left(\gO^2\right),&
\gthh &:= \gth + \gO \gth_1^A + O\left(\gO^2\right).
\end{align*}
Let us introduce
\begin{equation*} 
\begin{array}{llccl}
\ggh_{\mu\nu} = &\quad\; \gOh^0 &\Big( & 2(d\uh)_{\mu}(d\gOh)_{\nu} + \hh_{AB} \left(d\gthh^A\right)_{\mu}\left(d\gthh^B\right)_{\nu} &\Big) \\ \\

&+ \; \;\gOh^1 &\Big( &  -\; \gThh \; (du)_{\mu}(du)_{\nu} + \Ch_{AB} \left(d\gthh^A\right)_{\mu}\left(d\gthh^B\right)_{\nu} &\Big) + \;\; O\left(\gOh^2\right)
\end{array}
\end{equation*}	
a direct computation shows that if $\ggh_{\mu\nu} = \go^2 g_{\mu\nu}$ then we must have
\begin{align*}
\hh_{AB} = (\go_0)^2 h_{AB} + O\left(\gO\right),
\end{align*}
\begin{align*}
\ud_0 &= \go_0,  & u_1 &= -\tfrac{1}{2\go_0} (\covD u_0)^2, & \gth^A_1 &= -\tfrac{1}{\go_0} h^{AB} \covD_{B}u_0,
\end{align*}
\begin{align*}
\tfrac{1}{2}\Ch_{AB}\big|_0 = \go_0\Big(  \tfrac{1}{2}C_{AB} + \tfrac{1}{\go_0}\covD_A\covD_B u_0 - \tfrac{2}{(\go_0)^2}  \covD_{(A} \go_{0} \covD_{B)} u_0 +\tfrac{1}{(\go_0)^2}\left(\tfrac{\god_0}{\go_0} + \tfrac{1}{2}\gTh\right)\covD_{A}u_0 \covD_B u_0 \Big)\Big|_0 + O\left(\gO\right),
\end{align*}
\begin{align*}
\tfrac{1}{2}\hh^{AB}\Ch_{AB} = \tfrac{1}{\go_0}\Big( \tfrac{1}{\go_0} \covD^C \covD_C u_0 + \tfrac{d-2}{\go_0} \;\go_1 + \tfrac{d-4}{(\go_0)^2} \covD^C u_0 \covD_C \go_0 -\tfrac{d-4}{2(\go_0)^2} (\covD u_0)^2 \left( \tfrac{\god_0}{\go_0} +\tfrac{\gTh}{2}\right) \Big) + O\left(\gO\right).
\end{align*}
If $\left(\uh , \gOh= \go \gO, \gthh\right)$ are BMS coordinates then $\hh^{AB}\Ch_{AB}=0$ i.e.
	\begin{align*}
	\go_1 = -\tfrac{1}{d-2} \covD^C \covD_C u_0 
	- \tfrac{d-4}{d-2} \tfrac{1}{\go_0} \covD^C u_0 \covD_C \go_0  + \tfrac{d-4}{d-2} \tfrac{1}{2 \go_0} (\covD u_0)^2 \left( \tfrac{\god_0}{\go_0} + \tfrac{\gTh}{2}\right).
\end{align*}

\subsection*{Acknowledgments}
The author was supported by a ``Chargé de recherche'' grant from FNRS. This project has also received funding from the European Research Council (ERC) under the European Union’s Horizon 2020 research and innovation programme (grant agreement No 101002551).

\printbibliography
\end{document}